\documentclass[aps,pra,twocolumn,superscriptaddress,10pt,floatfix,showpacs]{revtex4-1}
\usepackage[dvips]{graphics}
\usepackage{graphicx}
\usepackage{amsfonts}
\usepackage{amssymb}
\usepackage{amsmath}
\usepackage{subfigure}
\usepackage{color}

\begin{document}

\title{Fluctuating and dissipative dynamics of dark solitons in quasi-condensates}

\author{S.P. Cockburn}
\affiliation{School of Mathematics and Statistics, Newcastle University,
Newcastle upon Tyne, NE1 7RU, United Kingdom}

\author{H.E. Nistazakis}
\affiliation{Department of Physics, University of Athens, Panepistimiopolis, Zografos,
Athens 15784, Greece }

\author{T.P.\ Horikis}
\affiliation{Department of Mathematics, University of Ioannina, 45110 Ioannina, Greece}

\author{P.G.\ Kevrekidis}
\affiliation{Department of Mathematics and Statistics, University of Massachusetts,
Amherst MA 01003-4515, USA}

\author{N.P.\ Proukakis}
\affiliation{School of Mathematics and Statistics, Newcastle University,
Newcastle upon Tyne, NE1 7RU, United Kingdom}

\author{D.J.\ Frantzeskakis}
\affiliation{Department of Physics, University of Athens, Panepistimiopolis, Zografos,
Athens 15784, Greece }

\begin{abstract}
The fluctuating and dissipative dynamics of matter-wave dark solitons within harmonically trapped, partially condensed Bose gases is studied both
numerically and analytically. A study of the stochastic Gross-Pitaevskii
equation, which correctly accounts for density and phase fluctuations at finite
temperatures, reveals dark soliton decay times to be lognormally distributed at
each temperature, thereby characterizing the previously predicted long lived soliton trajectories within each ensemble of numerical realizations (S.P. Cockburn {\it et al.}, Phys. Rev. Lett. 104, 174101 (2010)). 
Expectation values for the average soliton lifetimes extracted from these distributions 
are found to agree well with both numerical and 
analytic predictions based upon the dissipative Gross-Pitaevskii model (with the same {\it ab initio} damping). 
Probing the regime for which $0.8 k_{B}T < \mu < 1.6 k_{B}T$, 
we find average soliton lifetimes to scale with 
temperature as $\tau\sim T^{-4}$, in agreement with predictions previously made for the 
low-temperature regime $k_{B}T\ll\mu$.
The model is also shown to capture the experimentally-relevant decrease in the visibility of 
an oscillating soliton due to the presence of background fluctuations.
\end{abstract}

\maketitle

\section{Introduction}

As intriguing realizations of quantum objects on a macroscopic scale, 
the existence of solitons within atomic Bose-Einstein condensates (BECs)
has led to much experimental and theoretical work. Both dark \cite{Burger1999,Dobrek1999,
Denschlag2000,Anderson2001,Becker2008,Weller2008,Shomroni2009,Dries2010}
and bright solitons \cite{Carr2000,Khaykovich2002,Strecker2002,Cornish2006}
have been observed within single species BECs. The existence of each can be argued as being linked to the robustness
of the system's nonlinear waves and ultimately with the system's (near-) integrabilty. This is chiefly responsible for the absence of dissipation,
for example, during soliton-soliton collisions \cite{Weller2008}.
For a decay mechanism to manifest, a substantial departure from the integrable limit must be enforced.

BEC experiments typically take place within an effectively harmonic three-dimensional (3D) trapping potential, the introduction of which
already lifts the integrability of the system. In principle, this means that solitons are no longer protected from decay by the infinite number of conservation laws, as in the integrable homogeneous 1D Gross-Pitaevskii equation (GPE).
In order that solitons be rendered dynamically stable against decay induced by
transverse excitations \cite{Anderson2001},
they should be produced in highly elongated gases \cite{Muryshev2002}, in which phase fluctuations are known to play an enhanced role \cite{Petrov2000}.
Then, in the absence of a thermal cloud, they are found to be stable in the special case of a longitudinal harmonic confining potential \cite{Busch2000b},
which has been shown to be a direct consequence of the periodic emission and
reabsorbtion of sound waves as the solitons oscillate in a harmonic potential
\cite{Parker2003,Parker2010}.

The first successful observations of dark solitons in BECs were made by Burger {\it et al.} \cite{Burger1999} 
in a somewhat elongated set-up, although the solitons were found
to decay rather rapidly upon reaching the condensate edge, an effect attributed
to thermal decay \cite{Muryshev2002,Jackson2007}.
More recent experiments have produced solitons with much longer lifetimes allowing for the observation of head-on collisions between solitons \cite{Weller2008,Stellmer2008}
and the dynamics of one or more soliton oscillations \cite{Becker2008,Weller2008}.
\begin{figure}[b!]
  \begin{center}
		\leavevmode
		\hspace{-0.45cm}\includegraphics[scale=0.45,clip]{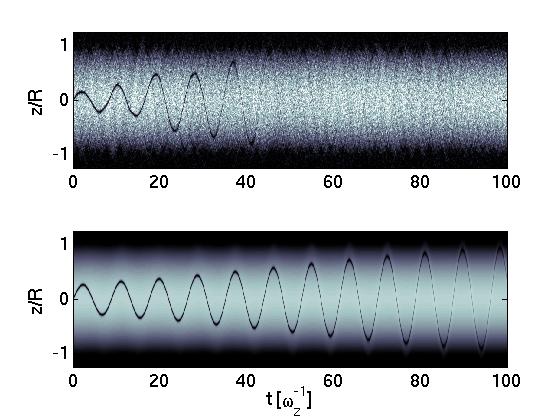}
		\vspace{-0.8cm}
  \end{center}
\hspace{2cm}
  \caption[]{Illustrative dissipative dark soliton dynamics in the presence (upper plot) 
						or absence (lower plot) of stochastically-sampled fluctuations for 
						$T \approx 1.9 T_{\phi} \approx 0.5 T_{\rm qc}$;
						Dissipative effects lead to the decay of the soliton in each case, while fluctuations (in the
						stochastic simulations)
						additionally lead to 
reduced soliton visibility and less regular oscillations.
    }
    \label{fig:front}
\end{figure}
In Ref. \cite{Becker2008}, dark solitons were created 
by a phase imprinting technique 
and found to exist for very long times,
with a clear oscillatory trajectory of the density notch
visible in the experimental data.
Interestingly, a strong shot-to-shot variation in soliton lifetimes
was also reported in this work: solitons were found to exist in single
realizations for times an order of magnitude longer than an average
trajectory was obtainable experimentally \cite{Becker2008,Weller_thesis}.
While attributed to small preparation errors in this experiment,
phase fluctuations are typically important in quasi one-dimensional (1D) systems,
suggesting that a similar effect might be expected to occur when introducing a
soliton within a phase fluctuating condensate \cite{Cockburn2010,Martin2010}.

Given the competition between fluctuations and dissipation
observed in experiments on dark solitons,
in this paper we perform a detailed numerical and analytical study
of the effect of each of these on the motion of dark
solitons within harmonically trapped,
phase fluctuating BECs.
Figure~\ref{fig:front} shows an example of the stochastic (upper panel)
and dissipative (lower panel) dynamics which we observe via the
theories we analyze here, discussed in Sec. II.

\section{Theoretical approach}

Given the particle-like behavior already
observed for dark solitons experimentally \cite{Stellmer2008,Weller2008,Becker2008},
the notion of a `heavy' soliton oscillating within 
a background of lighter thermal particles 
makes it tempting to draw an analogy to the Brownian motion of a particle moving within a fluid
of lighter particles, undergoing many scattering events.
The interaction between a dark soliton and thermal excitations in a BEC was
considered in \cite{Fedichev1999} by means of a kinetic equation, 
which was found to be of Fokker-Planck form under the assumption that
the momentum transfer per soliton-excitation interaction 
is much smaller than the soliton momentum.
In this work, we adopt a complementary Langevin approach which should thus 
also be well suited to the study of such systems, having been originally 
conceived for just this purpose \cite{Einstein1905,Langevin1908},
and we now discuss the pertinent model for weakly-interacting gases,
known as the stochastic Gross-Pitaevskii equation (SGPE).

\subsection{Stochastic Gross-Pitaevskii equation}

Since highly elongated trapping potentials are necessary if a soliton is
to be stable against transverse instabilities,
the inclusion of phase and density 
fluctuations is likely to prove essential in capturing all the salient
aspects of such necessarily ``low dimensional'' systems.
The 
SGPE \cite{Stoof1999,Stoof2001,Gardiner2003},
which we choose to employ here, is a model well suited to this task
as it captures both the dissipative and
fluctuating dynamics inherent to finite
temperature BECs, while additionally satisfying the required
balance between these two factors.

Within this scheme, the low-energy modes of our
effectively 1D 
system may be represented by the stochastic differential equation
\begin{equation}
  \begin{split}
    i\hbar\frac{\partial \Psi(z,t)}{\partial t}& = (1-i\gamma(z,t))
    \bigg[-\frac{\hbar^{2}}{2m}\frac{\partial^{2}}{\partial z^{2}}+V(z)\\
    &+g|\Psi|^{2}-\mu\bigg]\Psi(z,t)+\eta(z,t),
  \end{split}
  \label{eq:1dSGPE}
\end{equation}
where $\Psi(z,t)$ is a complex parameter describing the occupation of such low-lying modes, $V(z)=(1/2)m\omega_{z}^{2}z^{2}$
is the axial trapping potential, $g=2\hbar\omega_{\perp}a$
is the 
effective 1D interaction strength \cite{Olshanii1998} (with $a$ the s-wave scattering length),
and $\eta(z,t)$ is a complex Gaussian noise term, with correlations given by the relation
$\langle \eta^{*}(z,t) \eta(z',t') \rangle =2\hbar\gamma(z,t)k_{B}T\delta(z-z')\delta(t-t')$. The strength of the noise, and damping, due to contact with higher energy
thermal modes, is given by $\gamma(z,t)$.
\begin{figure}[tb!]
	\begin{center}
    \includegraphics[scale=0.3,clip]{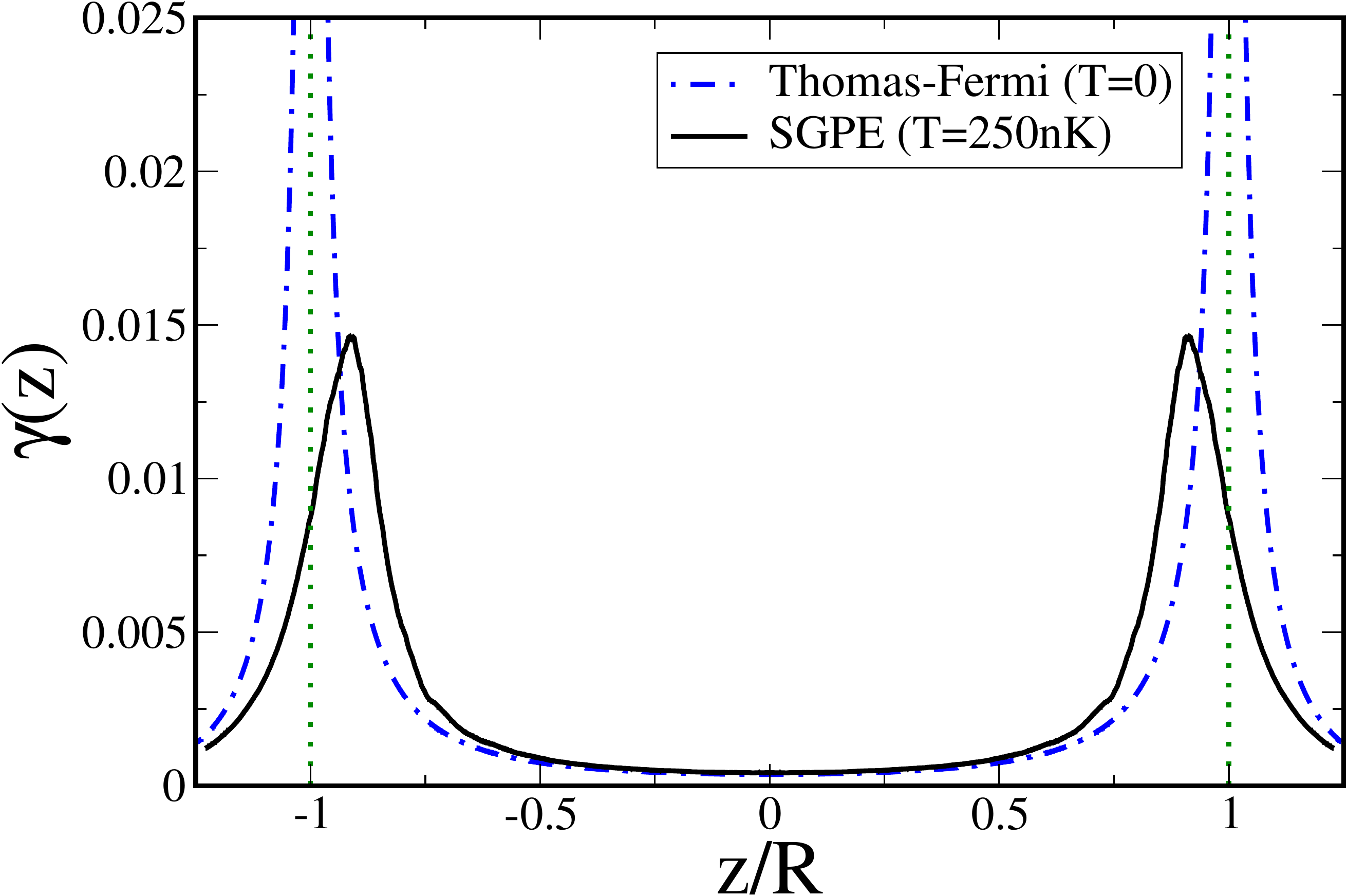}
	  \caption[]{
    (Color online) Form of the equilibrium dissipation $\gamma(z)$ at $T=250$nK when
		the SGPE density (black solid) and the Thomas-Fermi density (blue dot-dashed) are used
		in the mean-field potential $U(z)$.
		The vertical dotted lines indicate the $T=0$ 
        Thomas-Fermi radius, $R$.    
	  }
	\label{fig:T250nKgam}
  \end{center}	
\end{figure}

While the soliton will reside in the low-lying modes,
in arriving at Eq.~\eqref{eq:1dSGPE} it is assumed
that the high energy thermal atoms in the system
may be treated as though at equilibrium;
this should be valid for small perturbations, 
e.g. when introducing a dark soliton into a large system ($R \gg \xi$, where
$\xi = \hbar/\sqrt{m n g}$ essentially sets the soliton width)
which would have little effect on the high-lying modes,
which should instead remain close to equilibrium.
In this picture, the high- and low-energy systems are assumed to be in diffusive and
thermal equilibrium with a common temperature $T$ and chemical potential $\mu$.
Moreover, due to the large mode occupations typical within degenerate Bose gases,
we are well justified in the further assumption that the low-energy modes
are highly occupied for the classical
field approximation to be valid
\cite{Svistunov1991,Stoof2001,Duine2001,Davis2001,Sinatra2001,Goral2001}.

\subsection{Form of Dissipation}

Within the formulation of Stoof \cite{Stoof1999,Stoof2001}, 
the quantity which parametrizes the strength of
the noise and damping is the so-called Keldysh self-energy $\hbar\Sigma^{K}(z,t)$, which
may be related to the dissipation term in Eq.~\eqref{eq:1dSGPE} via
$\gamma(z,t)=i\beta\hbar\Sigma^{K}(z,t)/4$ \cite{Stoof2001,Duine2001}.
Following the methods of Refs.~\cite{Bijlsma2000,Duine2001}, and assuming the
system to be close to equilibrium, the integral to be
evaluated for the dissipation is
\begin{equation}
	\begin{split}
		\gamma(z)=\frac{4\beta m a^2}{\pi\hbar^{2}}
		\int d{E}_{2} \int d{E}_{3}\ S({k}_{1},{k}_{2},{k}_{3})
		(1+{N}_{1}){N}_{2}{N}_{3}
	\end{split}	
	\label{eq:gamma_int}
\end{equation}
where $0<E_{i}<\infty$, $N_{i}=[\exp(\beta(E_{i}+U(z)-\mu))-1]^{-1}$
is a Bose-Einstein distribution,
$U(z)=V(z)+2g\langle |\Psi(z)|^2 \rangle$, $E_{1}=E_{2}+E_{3}+U(z)-\mu$, and
\begin{equation}
	\begin{split}
S(k_{1},k_{2},k_{3})=&\frac{1}{2}\left\{\textrm{sgn}(k_{1}+k_{2}-k_{3})+\textrm{sgn}(k_{1}-k_{2}+k_{3})\right.\\
&\left.-\textrm{sgn}(k_{1}+k_{2}+k_{3})-\textrm{sgn}(k_{1}-k_{2}-k_{3})\right\},
	\end{split}
\end{equation}
with $k_{i}=\sqrt{2 m E_{i}/\hbar^2}$.

The damping may be evaluated numerically (e.g. using Gauss-Legendre quadrature), 
an example of which is shown in Fig.~\ref{fig:T250nKgam}. The data shown
is for $T=250$nK and illustrates two cases, when the density in the 
potential $U(z)$ is given by the SGPE density (black solid line)
or the $T=0$ Thomas-Fermi density (blue dot-dashed line)
[cf. Fig.~2 of Ref.~\cite{Duine2001}].
A notable feature in the form of $\gamma(z)$ is the peak close
to the Thomas-Fermi radius, which indicates that the scattering rate
is largest near the edge of the quasi-condensate.
Also, the peaks in $\gamma(z)$ calculated using the
SGPE density (black solid curve)
are noticeably closer to the trap center than for the
$T=0$ Thomas-Fermi case at the same $\mu$ (blue dot-dashed curve),
due to the effects of thermal depletion on the equilibrium density profile. 
We also note that the shape of $\gamma(z)$ is qualitatively similar to the 
spatial form of the scattering
rates found within numerical implementations of the Zaremba-Nikuni-Griffin
(ZNG) approach \cite{Jackson2002,ZNGbook}.

\subsection{Reduction to a dissipative Gross-Pitaevskii equation}

Neglecting the noise term of Eq.~\eqref{eq:1dSGPE} leads to
the following dissipative GPE (DGPE)
for the {\em condensate} wavefunction
(see e.g. the review of Ref.~\cite{Proukakis2008}):
\begin{equation}
  \begin{split}
    i\hbar\frac{\partial \phi(z,t)}{\partial t}& = (1-i\gamma(z,t))
    \bigg[-\frac{\hbar^{2}}{2m}\frac{\partial^{2}}{\partial z^{2}}+V(z)\\
    &+g|\phi|^{2}-\mu\bigg]\phi(z,t).
  \end{split}
  \label{eq:1dDGPE}
\end{equation}
This is expected to be a good model when damping effects are dominant over diffusion,
and representative of the mean field soliton dynamics in this case.
An advantage of deriving this model from the SGPE is that the damping
may be calculated in an {\it ab initio} manner, via Eq.~\eqref{eq:gamma_int},
rather than selected phenomenologically \cite{Choi1998,Tsubota2002}.

\subsection{Overview of numerical procedure}
\label{sec:numerics}

Before studying the soliton dynamics, an important preliminary step
is to generate a suitable equilibrium state. This state
differs between the DGPE and SGPE approaches: 
for the SGPE, the initial condition (the noisy field denoted by $\Psi$
in Eq.~\eqref{eq:1dSGPE}) 
is dynamically obtained at each temperature by equilibration with
the higher energy heat bath atoms - see, e.g., 
Ref.~\cite{Stoof2001,Proukakis2006a,Cockburn2009,Cockburn2011}.
For the DGPE, we instead use the
ground state of the GPE (the complex field denoted by $\phi$), which is
obtained efficiently by imaginary time propagation of the GPE.
We make the latter choice because the imaginary time solution to the GPE
is also the ground state of the DGPE, the action of which
is to drive the system towards this state, at a rate related to $\gamma$.
If instead we were to start with the SGPE initial condition as an input to the DGPE,
this would mean the system was already out of equilibrium, even prior to
introducing a perturbation such as a dark soliton.

To generate a dark soliton in {\it both} stochastic {\it and} purely dissipative simulations,
we multiply the equilibrium state by the soliton wavefunction
\begin{equation}
\psi_{\rm sol}(z)=\zeta {\rm tanh} (\zeta z / \xi) + i(v/c),
\label{eq:soli_psi}
\end{equation}
where $\zeta=\sqrt{1-(v/c)^2}$,
$\xi$ is the healing length, and $c$ is the speed of sound.
The input soliton velocity, $v_{\rm input}$, is chosen here such that $|v_{\rm input}|=0.25c$
in order to generate a relatively deep density notch (which is thus clearly identifiable over the thermal background noise).
We focus on this method of initial state preparation, rather than
full phase imprinting, as we wish to highlight the role of both the {\it initial} and
{\it dynamical} thermal noise even for a completely reproducible
preparation mechanism.

As $\phi(z)$ of Eq.~(\ref{eq:1dDGPE}) is a mean field, only one realization
of the DGPE is required at each temperature.
For the SGPE simulations, we must instead generate an ensemble of initial
states, $\{\Psi\}$, consisting of several hundred realizations, 
which we initially propagate to equilibrium with the higher energy thermal cloud. 
The equilibrium state is parametrized by the chemical potential, $\mu$, and temperature,
$T$, and both $\mu$ and $\gamma(z; \mu, $T$)$
remain the same between the SGPE and DGPE simulations
(the temperature dependence in the DGPE arises only through $\gamma$,
while the strength of the noise in the SGPE is also temperature dependent).

We choose to work with realistic trapping frequencies $\omega_{z}=2\pi\times10$~Hz and
$\omega_{\perp}=2\pi\times2500$~Hz, and set $\mu=395\hbar\omega_{z}$, which
corresponds to around $20000$ $^{87}$Rb atoms (at $T=0$).
To strike a balance between practical simulations and interesting dynamics due to
the interplay between dissipation and fluctuations,
we focus here on a temperature range $T=150$-$300$nK, which 
corresponds to $0.16<T/T_{\rm qc}<0.34$ 
and $6<T/T_{\phi}<13$,
where $T_{\rm qc}=N(\hbar\omega_{z})^2/\ln(2N)k_{B}$ \cite{Ketterle1996} and 
$T_{\phi}=N(\hbar\omega_{z})^2/\mu k_{B}$ \cite{Petrov2000}
respectively define `characteristic' temperatures for the onset of density and phase fluctuations.
Hence we deal with a partially condensed Bose gas, which is well within
the phase fluctuating regime.

\section{Analysis of stochastic dynamics}

The link between the quantization of classical theories
permitting soliton solutions, and dissipative quantum systems
was discussed in Refs.~\cite{Neto1991,Neto1992,Neto1993}.
There, it was highlighted that just as
the motion of a classical particle within a viscous environment has both a damped and
fluctuating component, the situation is the same in the quantum case.
The motion of such a classical particle can then
be characterized by two system properties, a damping due to the systematic
force applied to the particle and a diffusion related to this interaction \cite{Reif}.
This was shown to be true also in the quantum case \cite{Caldeira1983},
in which the dissipation manifests instead as a damping of the particle
wavepacket center of motion, whereas diffusive effects lead
to a spreading of the wave function for the particle \cite{Neto1993}.
For a soliton, the former would lead to a damping
of the motion of the soliton center, while the latter would give rise 
to an increased uncertainty in the soliton position.

For soliton solutions to integrable, classical one-dimensional theories,
in which case dissipation has no role,
the propagation in space is undamped and the soliton
dynamics is essentially captured by knowledge of the soliton center.
However, in the quantized field-theory at finite temperature,
not all degrees of freedom ``collaborate'' in the formation
of the soliton \cite{Neto1993} and the result is a
residual interaction which is shown to lead to a Brownian type motion of the soliton.
Therefore, the dynamics is no longer entirely captured by knowledge of
the center of mass alone, as the diffusive nature also becomes important.

In Ref.~\cite{Neto1993}, the Brownian nature of the soliton
motion is related to the coupling of the soliton to the other
system modes and excitations due to the presence of the soliton.
Damping and diffusion in the quantum
dissipative system must be temperature dependent, since the excitations
which scatter from the soliton are thermally activated.
Returning to the BEC context, 
there is an obvious analogy between this work and a
soliton propagating within a finite-temperature BEC.

Studies of 
weakly-interacting,
1D homogeneous Bose gases,
considering the purely dissipative dynamics,
found solitons to decay with a lifetime which varies with temperature
as $\tau \sim T^{-4}$ for $k_{B}T\ll\mu$ \cite{Fedichev1999,Gangardt2010}.
For $k_{B}T\gg\mu$, it was predicted that $\tau \sim T^{-1}$.
(This temperature dependence was also found
in a study on polarons \cite{Neto1992},
and in the more general case of the ``quantum impurity problem'',
applied in the setting of a heavy particle 
within a Luttinger liquid \cite{Neto1996}.)
The parameters in the present study are chosen such that 
$0.8\mu<k_{B}T<1.6\mu$, meaning we probe an
intermediate temperature regime which is hard to treat analytically
yet in which soliton decay can occur on a convenient time scale.
Such a regime has in fact been reached in a number of experiments on an atom chip, 
see e.g. Ref.~\cite{Armijo2010} and references therein, 
although this type of set-up has yet to
be used to observe the interesting soliton dynamics anticipated in such systems.
Despite this, several recent experiments 
have nonetheless provided motivation for modeling beyond purely
deterministic dissipative dynamics, with solitons observed 
to exist for times much longer than a
reproducible average trajectory could
be produced \cite{Becker2008}; indeed
preliminary work on thermal decay with $0.1\mu<k_{B}T<0.5\mu$ already 
found a significant spread in single trajectories \cite{Weller_thesis},
and such an effect is expected to be amplified as
fluctuations become more important.
The necessity for repeated runs in the SGPE formalism also 
has a strong link to the experimental approach of repeatedly creating 
successive BECs, so incorporates this effect naturally.

Finally, we note that Ref.~\cite{Gangardt2010} discussed the fact that
a full treatment of the soliton dynamics should
satisfy a fluctuation-dissipation theorem, which is not the case
in retaining only the damping aspects of the thermal background.
Fluctuations however complicate the analysis of experiments and the
stochastic simulations we have undertaken, as discussed in the
following Section. 

\subsection{Extracting soliton information}

\begin{figure}[b!]
  \centerline{
    \includegraphics[angle=0,scale=0.32,clip]{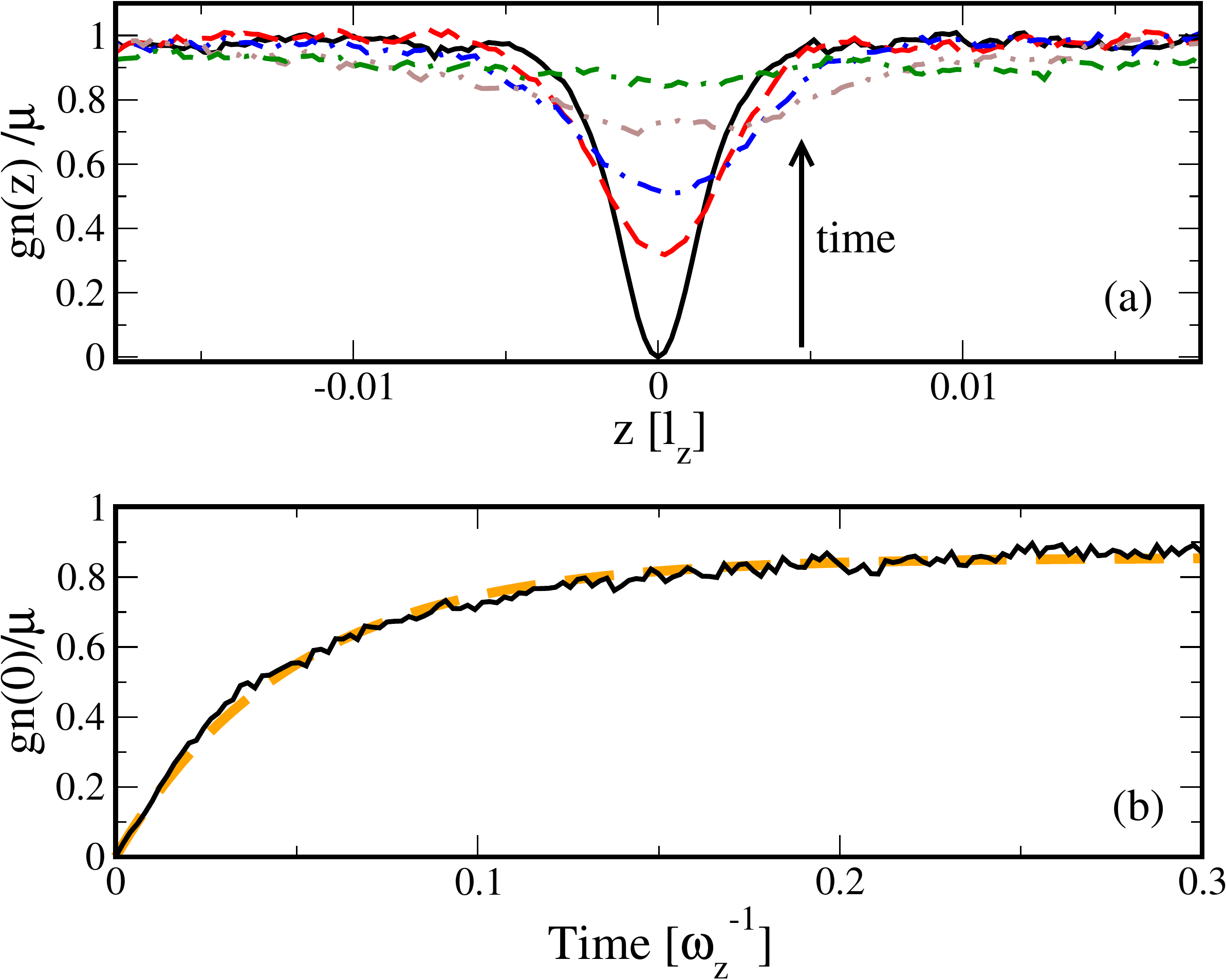}
  }
  \caption[]{
    (Color online) (a) Average density profile from an ensemble of stochastic simulations,
		with each realization containing an initially static, black soliton.
		Thermal	diffusion of individual solitons leads to the initially very deep
		notch {\emph average} density becoming gradually filled.
		Snapshots are shown at several times between the initial state (black solid line) to the final state
	  (dot-dashed green line). (b) Growth of the average central density tends towards the equilibrium value,
		as the soliton is lost in the average density profile,
		even though the solitons are still present within individual runs (see Fig.~\ref{fig:moving_sol}).
	}
  \label{fig:static_soli}
\end{figure}
Within the SGPE approach, observables are 
obtained by sufficient averaging
over many stochastic realizations. For example, the density
for an average over $N$ noise realizations,
${\Psi_{1},\Psi_{2},\ldots,\Psi_{N}}\equiv\{\Psi\}_{N}$ is given by
$\langle n(z)\rangle=\langle\Psi^{*}(z)\Psi(z)\rangle=(1/N)\sum_{i=1}^{N}\Psi_{i}^{*}(z)\Psi_{i}(z)$.
In order to extract information on the soliton present within
each stochastic realization,
we find we must extract the soliton trajectory prior to performing any averaging.
If we instead simply average over the set of density profiles to obtain $\langle n(z) \rangle$,
the soliton is quickly found to be ``washed out'',
despite the fact that a soliton is still present within each
individual realization $\Psi_{1},\Psi_{2},\ldots,\Psi_{N}$
-- this may be understood from
the fact that after some time solitons within different realizations are likely
to be at different positions, so averaging
over many such density profiles quickly leads to a loss of
information on the individual soliton positions.

This effect is illustrated for the case of an initially static soliton in
Fig.~\ref{fig:static_soli}. The average density is plotted in Fig.~\ref{fig:static_soli}(a),
showing the `average soliton' filling up over time; the density at the center of the sample,
where the soliton is initially positioned, is shown in Fig.~\ref{fig:static_soli}(b)
and grows towards the equilibrium value, with a growth in time which
goes as $\propto(1-\exp(-\Gamma t))$, with $\Gamma(>0)$ the growth rate. 
The experimental analogue of this effect was discussed in Ref.~\cite{Weller_thesis},
where it was noted that it is indeed single-shot soliton runs that should be
analyzed to measure decay, rather than averaged images, as the soliton
contrast is quickly found to be smeared out in the latter. 
In addition, Dziarmaga analyzed the diffusion of a soliton due to quantum fluctuations \cite{Dziarmaga2004},
while here it is thermal fluctuations which
cause the diffusion process
(see also \cite{Mishmash2009,Cockburn2010,Cockburn_PhD,Martin2010,Dziarmaga2010,Mishmash2010reply}).
\begin{figure}[b!]
  \centerline{
    \includegraphics[angle=0,scale=0.32,clip]{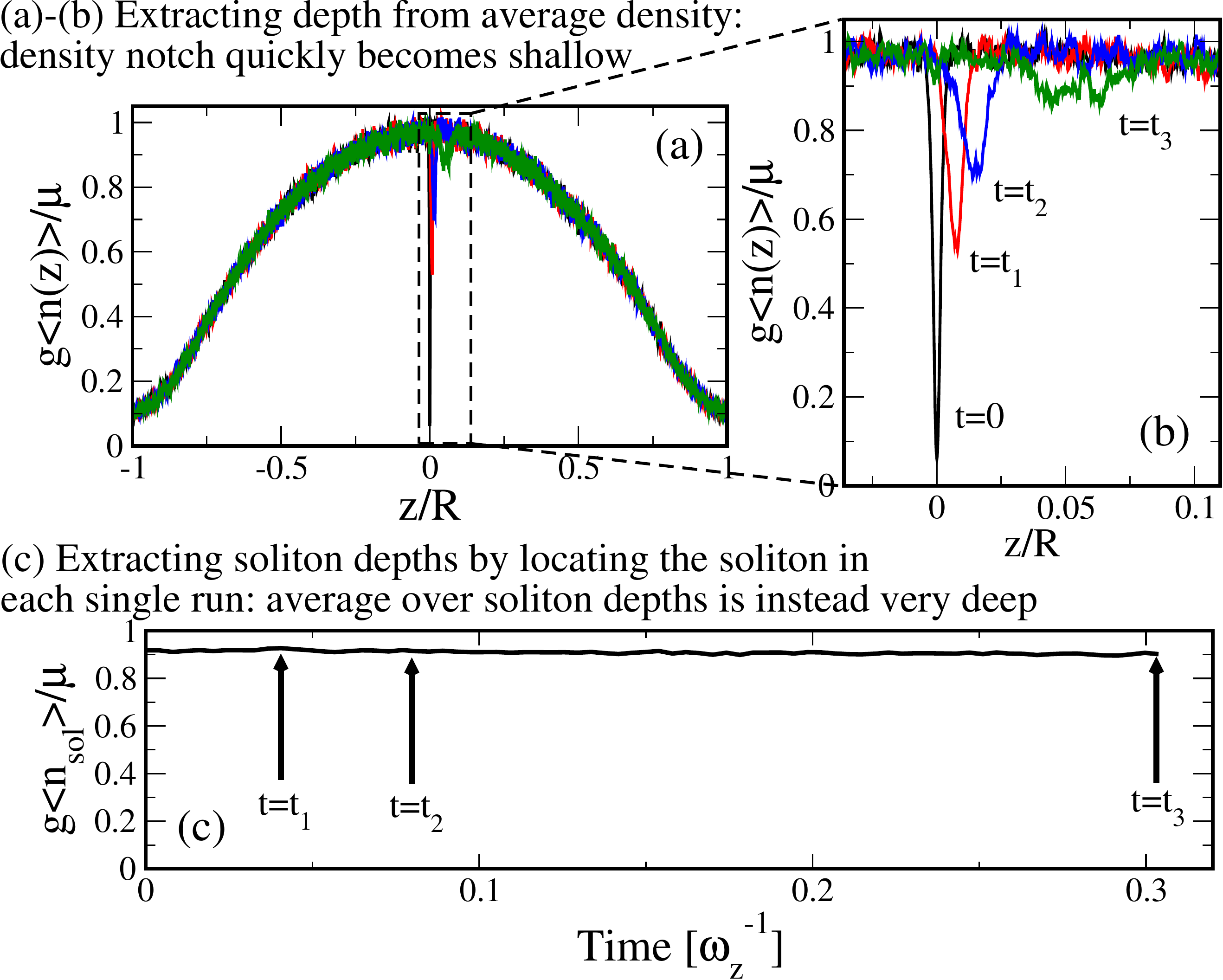}
  }
  \caption[]{
Typo in text in (c) within the figure 'soltion'.
(Color online) (a) Average density profile containing a moving soliton ($|v_{\rm input}|=0.25c$); (b) as
		in (a), but focused on the soliton region. (c) Average of the soliton depths extracted from each
		{\it individual} stochastic realization - this remains nearly constant meaning a 
		deep soliton remains within each individual run. Information on the soliton depths is lost in the 
		average {\it density} profile due to the soliton density notches of each run being, in general, 
		at different spatial points. This implies that soliton information should be extracted from each 
		realization, prior to any statistical analysis.
  }
  \label{fig:moving_sol}
\end{figure}

While soliton depth, or contrast, is quickly lost in average images due to statistical
effects on the motion, we now look at the fate of the solitons residing within
each single SGPE realization. This is illustrated in Fig.~\ref{fig:moving_sol},
where we instead consider a moving soliton. 
The soliton depth, $n_{\rm sol}$, is measured relative to the background density,
meaning, for example, $n_{\rm sol}$ equal to the background density would correspond to 
the static, black soliton of Fig.~\ref{fig:static_soli}.
As in the case of a static soliton,
the moving soliton is quickly lost in the average density profile, as shown by the density snapshots
of Figs.~\ref{fig:moving_sol}(a)-(b). However, if we look instead at Fig.~\ref{fig:moving_sol}(c),
then we see that each single stochastic realization in fact 
still contains a very deep soliton:
this plot shows the result of measuring $n_{\rm sol}$,
the depth of the soliton 
relative to the background density, within each {\em individual}
realization of the ensemble $\{\Psi\}_{N}$.
Fig.~\ref{fig:moving_sol}(c) shows the average of this set of depths,
$\langle n_{\rm sol} \rangle$.
This average depth is quite  different to that obtained by averaging over the individual density 
profiles (i.e. calculating $\langle\Psi^{*}(z)\Psi(z)\rangle$ as in Figs.~\ref{fig:moving_sol}(a)-(b)), 
as importantly the soliton is first located within a single realization (and 
this location generally varies from realization to realization) 
before information on its depth is extracted.
Simply averaging the density profiles of individual runs after some time
leads to a smooth profile, as the soliton density minimum 
in a particular run becomes outweighed by the higher number of runs in which there 
is no soliton at that particular spatial point. 
In our analysis, we therefore follow the 
density notch associated with the soliton present within each realization,
as found also to be experimentally most consistent when considering
soliton decay \cite{Weller_thesis}.

We can only track a soliton until the point where it becomes indistinguishable 
over the background density fluctuations which defines the soliton decay time
(in an analogous manner to experimental observations);
carrying this out for each individual realization allows
us to obtain an ensemble of soliton trajectories and so a
distribution of soliton dynamics $\{\Psi\}_{N}$ which we can then analyze.
This means of extracting data from the stochastic simulations appears
consistent with the idea of the soliton center as a
good quantum dynamical variable \cite{Neto1992}.

\subsection{Spread in initial soliton velocity}

As discussed in Section \ref{sec:numerics}, we introduce a soliton
of prescribed velocity ($|v_{\rm input}|=0.25c$) in an identical manner within
each simulation. Despite this, one might expect 
a spread in initial velocities due to the 
the fluctuating background density, $n=|\Psi|^2$, 
and the formula relating this to
the soltion velocity,
$v/c=\sqrt{1-n_{\rm sol}/n}$.
\begin{figure}[tb!]
  \centerline{
    \includegraphics[angle=0,scale=0.3,clip]{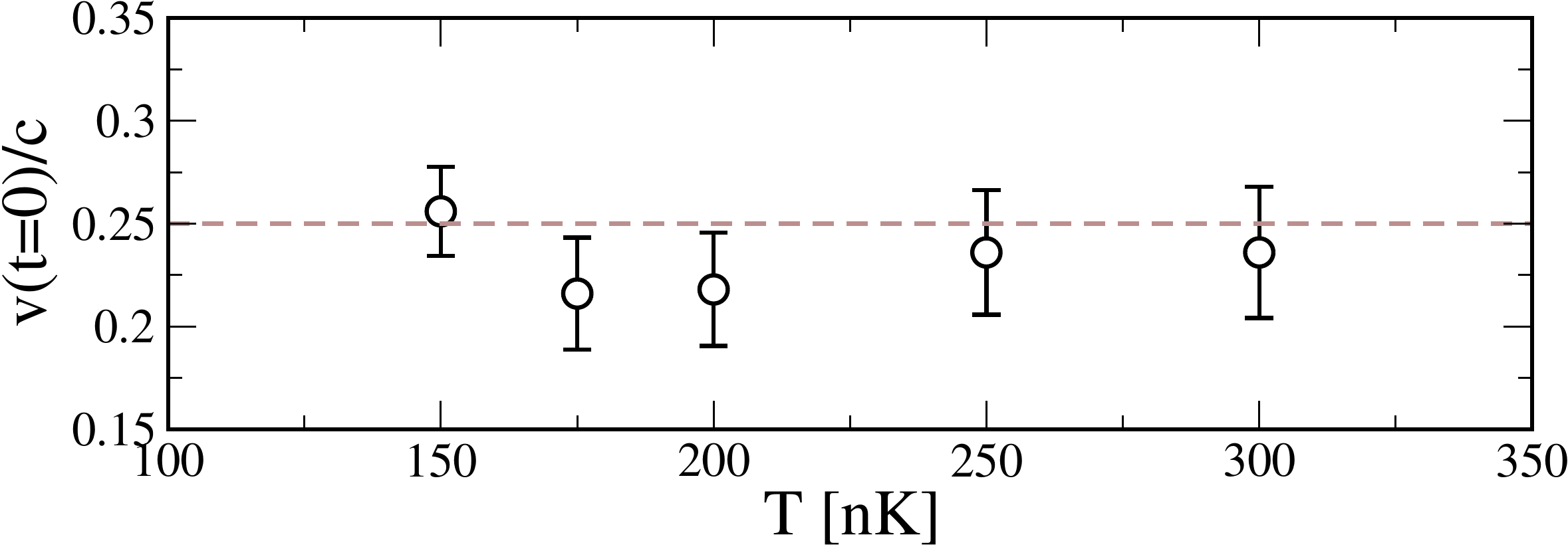}
  }
  \caption[]{(Color online) Comparison of the average measured soliton velocity (black circles)
						 to the input velocity (dashed brown line) for various temperatures.
						 The error bars indicate the standard deviation of the measured velocities.
  }
  \label{fig:init_speeds}
\end{figure}
Fig.~\ref{fig:init_speeds} shows the mean initial velocities from a
few hundred realizations
versus temperature, with error bars showing the standard deviation.
This plot shows a tendency for the soliton velocity to be lower 
than the prescribed value (shown by the dashed horizontal line)
for higher temperatures, and consequently larger fluctuations.
\begin{figure}[b!]
  \centerline{
    \includegraphics[angle=0,scale=0.320,clip]{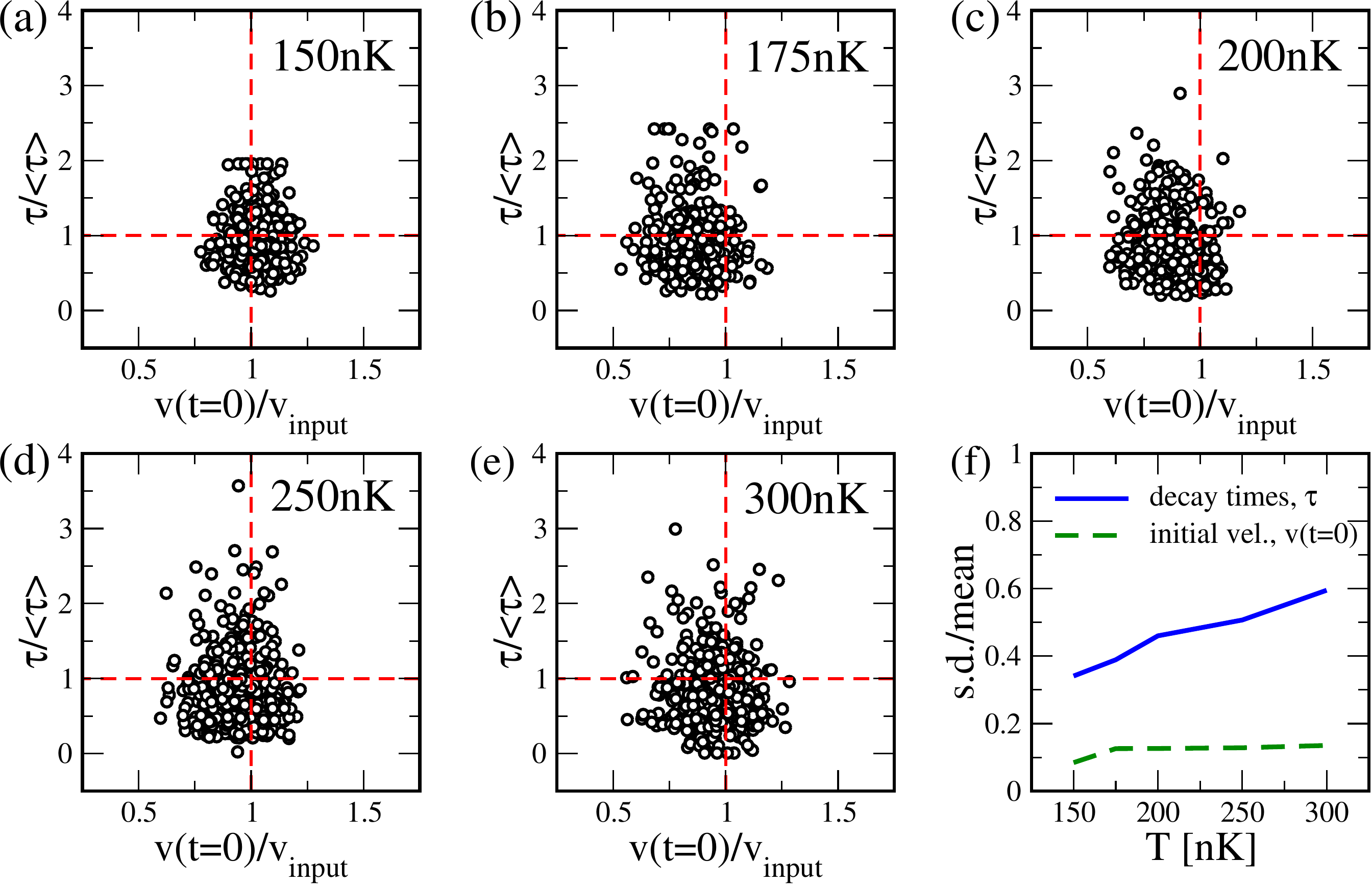}
  }
  \caption[]{(Color online) Scatter plots showing decay times, scaled to
						 the ensemble average (y-axis), vs. initial soliton velocity, scaled
						 to the input velocity (x-axis); $T=$ (a) $150$~nK, (b) $175$~nK, (c) $200$~nK
						 (d) $250$~nK and (e) $300$~nK. (f) shows the standard deviation in the
						 decay times (solid blue line) and initial velocities (dashed green line)
						 vs. $T$.
  }
  \label{fig:init_vel_vs_tau}
\end{figure}


\subsection{Influence of initial vs. dynamical noise}
\begin{figure*}[hbtp!]
  \centerline{
    \includegraphics[angle=0,scale=0.35,clip]{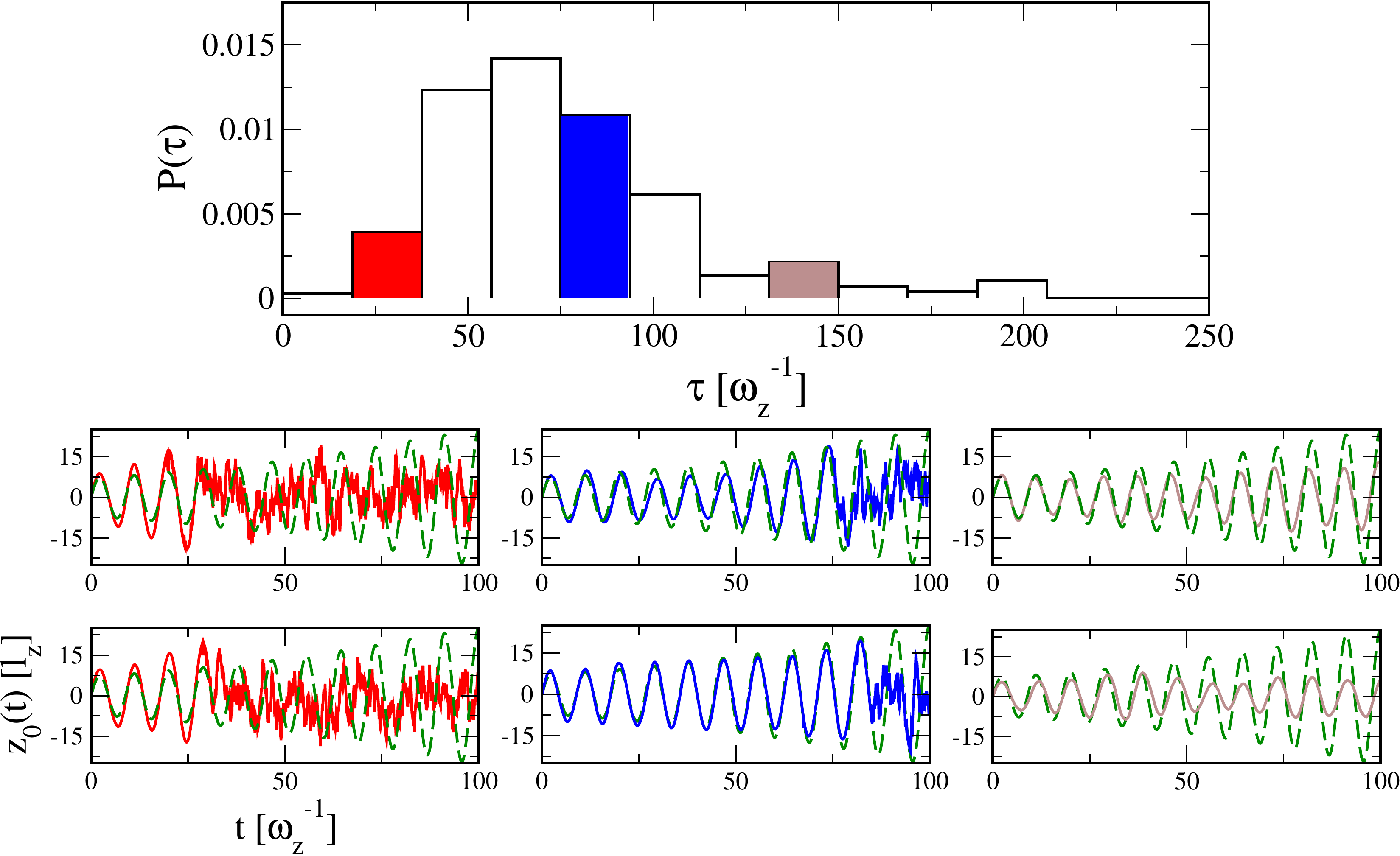}
  }
  \caption[]{(Color online) Histogram of stochastic soliton decay times (upper plot) and two representative sets of
						trajectories from the highlighted bins (lower plots).
						The solid lines in the lower plots show stochastic trajectories while the
						dashed green lines is the purely dissipative result. The left/middle/right most plots
						have decay times within the corresponding left/middle/right highlighted histogram bins.
						The trajectories from the middle highlighted
						bin have a decay time close to the mean value, and we find the trajectories to be
 						qualitatively similar to that of the dissipative case.
  }
  \label{fig:repre-traj}
\end{figure*}
In addition to a variation in initial velocities,
fluctuations due to the {\it dynamical} noise
term of the SGPE also affect the solitons throughout their lifetimes
(see e.g. Ref.~\cite{Duine2004} for related work on fluctuations during
vortex decay).
We use the decay time, i.e. the time at which our tracking algorithm
can no longer distinguish a soliton over the background noise, as an
observable to measure the soliton dynamics. 

To study the relative importance of initial versus dynamical noise,
we construct a scatter plot (see Fig.~\ref{fig:init_vel_vs_tau}(a)-(e )) 
showing, on the $x$-axis, the initial velocity, $v(t=0)$, scaled to
the input velocity used for deterministic soliton creation, $v_{\rm input}$.
On the $y$-axis, we show
the decay time, $\tau$, scaled to the ensemble average $\langle\tau\rangle$.
The vertical dashed red line indicates the velocity $v_{\rm input}$, 
while the horizontal dashed line indicates the ensemble
average decay time, $\langle\tau\rangle$.
At all but the lowest temperature of Fig.~\ref{fig:init_vel_vs_tau}(a),
there are more points to the left of the expected velocity
(vertical dashed line) than the right,
suggesting that the soliton generated within many of the realizations had
a velocity which was lower than the input velocity of Eq.~\eqref{eq:soli_psi}
(consistent with Fig. 5).

Looking instead at the vertical trend, we see that at each temperature
there are several solitons which exist for very long times relative to
the ensemble average (horizontal dashed line). Two examples of
such long-lived trajectories are shown in the lower, rightmost plots of
Fig.~\ref{fig:repre-traj}.

The data of Fig.~\ref{fig:init_vel_vs_tau} also allows us to indicate whether
it is the variation in initial velocities or the dynamical noise which
has greatest influence on the variation in soliton dynamics.
The relative variation due to each of these effects is quantified
in Fig.~\ref{fig:init_vel_vs_tau}(f) which shows the standard deviation
for each quantity, scaled to the relevant mean. 
It is clear that the relative spread in
the decay times is far greater than the spread in the initial
velocities, indicating that the dynamical noise has a larger cumulative influence
on the observed variation in the soliton dynamics.

Fig.~\ref{fig:repre-traj} summarizes the large variation in
soliton dynamics that is observed, despite the identical means of
soliton generation employed.
The top panel shows a histogram of soliton decay times, with a 
selection of representative soliton
trajectories in the panels below. 
The colour/position of the trajectories in the lower plots
correspond to the highlighted bins of the histogram.

In the case of the leftmost and middle plots, 
we see the trajectories are no longer oscillatory after some time,
and instead become noisy. This is the point at which the soliton is lost, and defines
the decay time (after which the graph simply represents numerical noise).
The middle plots show trajectories for solitons from
the bin containing the mean decay time ($5$th bin from left, highlighted blue),
which are typically very close to the
trajectory obtained from the DGPE simulation at this temperature.
Finally, in the rightmost plots, we consider a bin corresponding to
decay times which are longer than the ensemble mean;
their amplitudes are far below that
of the purely dissipative trajectory, shown by the dashed green line,
indicating that the noise can act in some cases to stabilize a soliton
against decay, relative to the purely dissipative evolution at
the same temperature.

\subsection{Distribution of soliton decay times}
\label{sec:distr}

\begin{figure}[t!]
  \centerline{
    \includegraphics[angle=0,scale=0.3,clip]{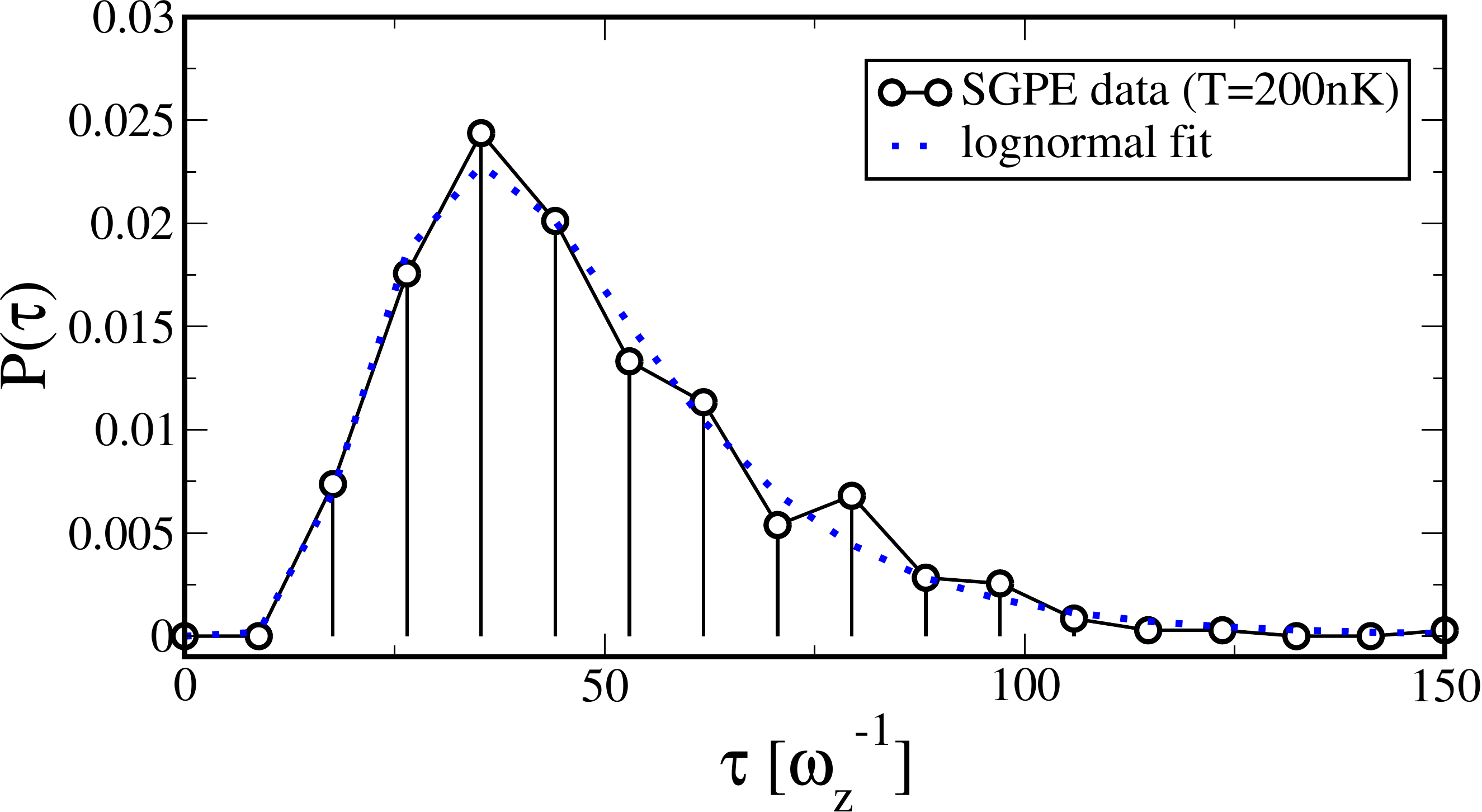}
  }
  \caption[]{
    (Color online) Histogram of soliton decay times for an ensemble of few hundred SGPE runs
		(solid black circles) and a fit to a lognormal distribution (dotted blue). 
  }
  \label{fig:lognormal_fit}
\end{figure}

An example of the distribution of decay times at $T=200$~nK
is shown in Fig.~\ref{fig:lognormal_fit}.
It is clear that the distribution of soliton decay times is non-Gaussian,
and we instead find it to be well fitted by a lognormal distribution,
\begin{equation}
P(\tau)=\frac{1}{\tau\sigma\sqrt{2\pi}}\exp\left[ \frac{-\left(\ln \tau - m\right)^{2}}{2\sigma^{2}} \right],
\label{lognorm}
\end{equation}
where $m$ is the mean and $\sigma$ the standard deviation of $\ln \tau$.
A lognormal model is often applied to a system in which
decay is caused by random events, which causes decay at
a rate proportional to the amount already present, so
in other words a runaway process. For example,
Kolmogorov suggested
that for anything which decays in this multiplicative way,
the time to failure should follow a lognormal distribution \cite{Kolmogorov1941}.

This distribution has a long tail at large decay times, which reflects
the finite probability of some long lived solitons within a given
ensemble of simulations.
Interestingly, extreme cases of long-lived solitons have been observed experimentally
within single runs, while the typical, reproducible decay time for the experiment
was around an order of magnitude lower \cite{Becker2008}. 
That soliton decay times follow a lognormal
distribution is a possible reason for this observed behavior.

It is also interesting to note that
a lognormally distributed soliton amplitude,
which we denote $S(t;\gamma,\mu)$,
would solve a stochastic equation of the form
\begin{equation}
	dS(t;\gamma,\mu) = m S(t;\gamma,\mu) dt + \sigma S(t;\gamma,\mu) dW,
\label{eq:discussion}
\end{equation}
which is the stochastic differential equation
defining a geometric Brownian motion (here, $m$ 
is the average growth rate,
$\sigma$ is the so-called volatility and $dW$ denotes a Weiner process).
As the soliton depth determines how far
from the trap center the turning point of the
motion is, it is therefore directly related to the amplitude of
the oscillations. In turn, as the decay time is determined by
reaching a certain depth, then we expect the distribution in
this variable to display a similar lognormal distribution.
This would imply that the amplitude of the soliton oscillations undergoes a
lognormal random walk. 

Eq.~\eqref{eq:discussion} should also be compared to the subcritical 
soliton equation of motion derived in Ref.\cite{Cockburn2010}, 
and also that discussed in Section~\ref{sec:analytics} [Eq.\eqref{sbcr}].
The latter two predict soliton oscillations with an 
exponentially increasing amplitude envelope dependent upon the damping,
and the solution to Eq.~\eqref{eq:discussion} would also correspond to such an exponential function
in the limit that $\sigma\rightarrow0$ (or, equivalently, fluctuations are neglected).
Conversely, fluctuations might be introduced retrospectively to the dissipative model
by allowing the damping to become a stochastic variable: 
a similar idea has been applied in describing the collisions
between optical solitons \cite{Chung2005}, where it was found that
collisions could be described by a nonlinear Schr\"{o}dinger equation
perturbed by stochastic parameters
obeying strongly non-Gaussian statistics. Interestingly,
in this case the soliton amplitude was also
found to be lognormally distributed.

\begin{figure}[b!]
  \centerline{
    \includegraphics[angle=0,scale=0.24,clip]{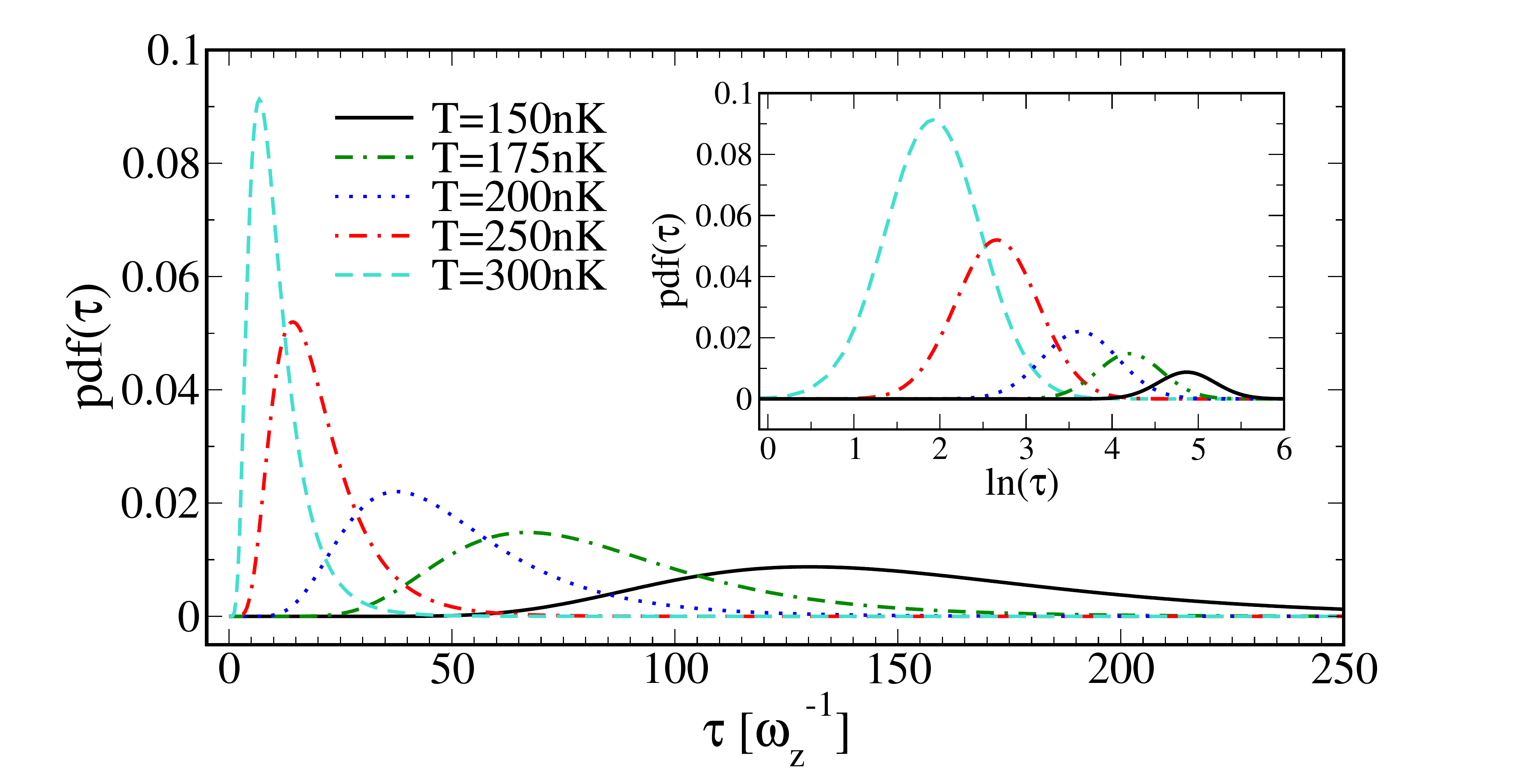}
  }
  \caption[]{
	  (Color online) Lognormal probability distributions obtained from fitting soliton decay time
		histograms obtained at the temperatures indicated. Inset: Same data with a log-scale x-axis;
		it is the logarithm of the decay time which is Gaussian distributed.
  }
  \label{fig:lognorm_fits}
\end{figure}
\begin{figure*}[htb!]
  \centerline{
    \includegraphics[scale=0.3,clip]{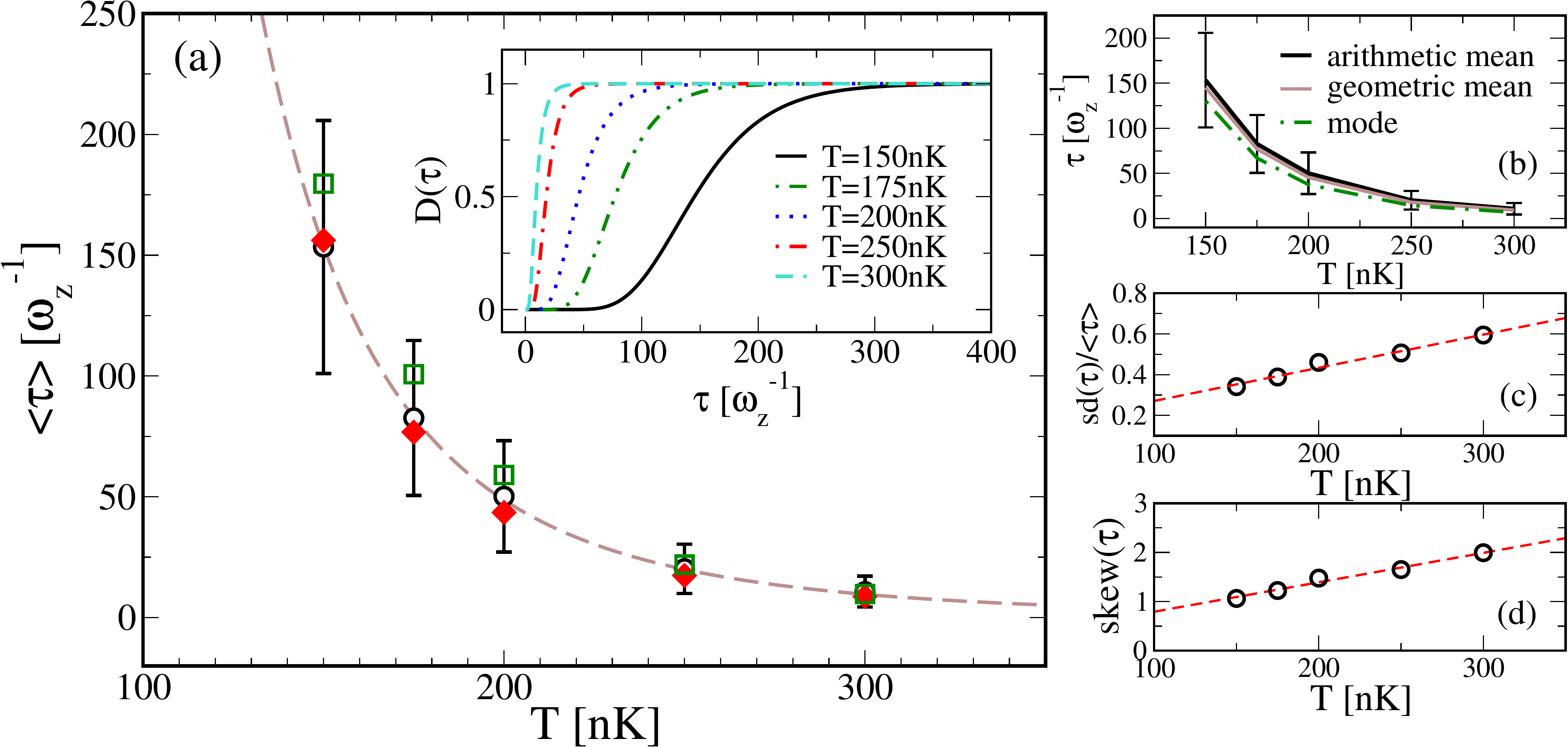}
  	}
  \caption[]{
    (Color online) Analysis of the decay time distributions: (a) Expectation value of the
		SGPE soliton decay time (black circles; error bars denote one standard deviation) versus
		temperature. Corresponding DGPE decay times are also shown (filled red diamonds) and a fit
		to a function $\propto T^{-4}$ (brown dashed line).
		Green squares show analytical predictions from the analysis of Section \ref{sec:analytics};
		the inset shows the cumulative probabilty
		distribution: the fraction of solitons which have decayed, versus time, for each temperature.
		(b) SGPE expectation value (solid black) compared to
		the geometric mean (brown solid line) and modal (green dot-dashed line) values for each
		temperature; (c) standard deviation and (d) skewness of $\tau$ versus temperature.
  }
  \label{fig:time_analysis}
\end{figure*}

\subsection{Decay times vs. $T$: Analyzing the distributions}

In order to extract meaningful quantities from our results, we must 
analyze the distributions of decay times which we obtain from the stochastic 
simulations. To do so, we proceed by fitting a normalized histogram of decay times
at each temperature to Eq.~\eqref{lognorm}. At each temperature, we find that the
fit matches the numerical data well, suggesting that the
underlying mechanism of the soliton decay is multiplicative in nature.
The results of fitting the decay times for all
temperatures considered are shown in Fig.~\ref{fig:lognorm_fits};
the inset shows the same data with the $x$-axis on a log-scale.
We can see from the main plot of Fig.~\ref{fig:lognorm_fits} that the
distributions of decay times become increasingly shifted towards
the origin with increasing temperature,
as dissipative effects reduce the soliton lifetimes.
The inset highlights that it is the logarithm of the decay times
which is Gaussian distributed. From the inset it is clear also
that the variance of $\ln(\tau)$ increases as temperature is increased.

An analysis of the behavior of the decay time distributions with temperature is
given in Fig.~\ref{fig:time_analysis}. To extract the average soliton 
behavior, we consider first the expectation value, $\langle\tau\rangle$, which for the 
lognormal distribution of Eq.~\eqref{lognorm}, is given by 
$\langle\tau\rangle=\exp{[\mu+\sigma^2/2]}$. The variance is instead
calculated as $\mathrm{Var}[\tau]=(\exp{[\sigma^2/2]}-1)\exp{[2\mu+\sigma^2]}$,
and we use $\sqrt{\mathrm{Var}[\tau]}$ to generate the error bars in the SGPE
data of Fig.~\ref{fig:time_analysis}(a).

Referring to Fig.~\ref{fig:time_analysis}(a), we 
find that $\langle \tau \rangle$ for the SGPE results (black circles)
closely follows a $T^{-4}$ behavior \cite{Cockburn_PhD}, shown by the brown dashed line.
Therefore the average behavior follows the same scaling predicted by
the models in \cite{Fedichev1999,Gangardt2010} for $k_{B}T\ll\mu$.
We work instead in the regime where
$k_{B}T\sim\mu$, which is difficult to treat
analytically, however find the low temperature result to
extend to this regime as well (although our numerics also
relies on the classical field approximation).

In order to compare meaningfully between the current SGPE analysis
and the DGPE soliton dynamics,
we must account for the changing level of background density fluctuations
as temperature is varied within the SGPE
\cite{Cockburn2010,Cockburn_PhD}. To do so we have measured the
minimum depth at which solitons can still be resolved in the SGPE runs,
at each temperature \footnote{The current analysis is based upon a time average of
readings from the soliton tracking routine for a system with no soliton, 
which therefore measures only background noise.
This differs slightly to the analysis of Ref.~\cite{Cockburn2010}, 
where the depth measurement was instead based 
on the average {\it maximum} value of the background noise.
The effect of this was a slight underestimation
of the DGPE decay times relative to the SGPE results in
our previous work.}.
This is then used to extract a soliton decay time within 
the DGPE simulations, i.e. the time for the soliton to decay to 
the temperature dependent depth extracted from the SGPE simulations. 
The numerical DGPE results are shown by the filled red diamonds in
Fig.~\ref{fig:time_analysis}(a), and display very good agreement with the
ensemble average stochastic results, $\langle\tau\rangle$. 
We additionally show the results of
our analytic model for the DGPE dynamics (hollow green squares), which is discussed 
in the following Section~\ref{sec:analytics}. The decay times obtained
within this model also agree well with the numerical DGPE and SGPE results,
when limits on the soliton visibility due to thermal fluctuations are accounted for
(see Section \ref{sec:soli_vis} for details).

As indicated above, the SGPE decay time
distributions are not symmetric (see Fig.~\ref{fig:lognorm_fits})
and feature a long tail at long soliton lifetimes for all temperatures considered.
Our results suggest also that the noise stabilizes a certain number of
solitons created within
each stochastic ensemble of realizations against decay,
relative to a soliton undergoing purely dissipative dynamics
under the DGPE. This was also apparent from the
decay time histograms shown in 
Refs.~\cite{Cockburn_PhD,Cockburn2010}.
The inset to Fig.~\ref{fig:time_analysis}(a) shows the cumulative distribution
function, $D(\tau)$, which measures the fraction of solitons which have decayed
at any time, based on the $P(\tau)$ curves of Fig.~\ref{fig:lognorm_fits}.
Clearly, dissipative effects are more dominant in the high temperature data,
with all solitons found to have decayed at relatively short times.
That the cumulative distribution function has a slow asymptote towards unity,
is another way to see that some fraction of the solitons live for far longer
than the average decay time. It is also interesting to note that scaling the 
time axis in each curve to the average at that temperature $\langle\tau\rangle$, 
we find each graph to collapse down close to a single curve.

Fig.~\ref{fig:time_analysis}(b) compares several
different measures of the distribution function:
the arithmetic mean, or $\langle\tau\rangle$,
the geometric mean and the modal value.
Comparing these, we see that the modal values
give consistently lower decay times than the expectation value, which is
a consequence of the peak in the distributions of Fig.~\ref{fig:lognorm_fits} being
located to the left of the mean.

In Fig.~\ref{fig:time_analysis}(c) we measure the standard deviation of $\tau$,
scaled to $\langle\tau\rangle$, which shows that the relative spread of the distributions
increases monotonically with $T$. This is as one might expect, given that higher
temperatures give rise to more fluctuations and so a wider variation between realizations.
This also shows that while the lowest temperature distribution of Fig.~\ref{fig:lognorm_fits}
appears far wider than the highest temperature case, 
the scaled standard deviation (s.d.) illustrates
that it is actually smaller.
Finally, Fig.~\ref{fig:time_analysis}(d) shows the skewness of the distribution functions,
which is also found to increase with temperature.

Next, we choose to decouple the effects of fluctuations and
damping, and consider
the dissipative behaviour alone through the mean field DGPE.
A crucial ingredient in both the analytic and numerical calculations which
we present, is the form of the damping term which we obtain from the SGPE formalism.

\section{Analytical approach for dissipative dynamics}
\label{sec:analytics}

We now consider the DGPE of Eq.~(\ref{eq:1dDGPE}).
In order to proceed with analytical calculations,
we first characterize the form of $\gamma(z)$ in terms
a simple analytical fitting function.
We find the dissipation $\gamma(z)$ to be
well approximated at all temperatures considered by the function
\begin{equation}
\gamma(z)=\frac{a}{(z+c)^2+d} + \frac{a}{(z-c)^2+d},
\label{eq:approx1}
\end{equation}
which is the sum of two Lorentzians, centred near the edge of the quasi-condensate.
The parameters $a$, $c$, and $d$ depend in general on temperature, $T$,
the trapping frequencies, chemical potential, and atomic species 
(although all except $T$ remain fixed in our analysis).
Various values of these parameters, indicating their temperature dependence,
are provided in Table~\ref{table_acd}.

\begin{table}[b]
\begin{tabular}{|c|c|c|c|}
\hline
T(nK) & a & c & d \\
\hline
125 & 0.0137 & 26.86 & 2.555 \\
\hline
150 & 0.0219 & 26.63 & 3.128 \\
\hline
175 & 0.0330 & 26.43 & 3.464 \\
\hline
200 & 0.0511 & 26.29 & 4.838 \\
\hline
250 & 0.1031 & 25.93 & 7.114 \\
\hline
300 & 0.1812 & 25.69 & 9.860 \\
\hline
\end{tabular}
\caption{Values of the parameters $a$, $c$ and $d$ [cf. Eq.~(\ref{eq:approx1})] for various values of temperature $T$.}
\label{table_acd}
\end{table}

We will make use of the form given by Eq.\eqref{eq:approx1} in the following Section,
in which we generalize our previous work on spatially constant dissipation reported in Ref.~\cite{Cockburn2010},
by developing analytic solutions for the more general case of spatially varying dissipation.

\subsection{Analytical approximations}
\label{sec:an_approx}

To perform our analytical work based on perturbation theory for dark solitons
we follow the ``routine'' procedure of rescaling the equation in appropriate
units. In particular, the 1D dissipative GPE (DGPE) model may also be written
in the following dimensionless form,
\begin{eqnarray}
[i-\gamma(z)] \partial_{t}\psi = \left[\frac{1}{2} \partial_{z}^{2}
+ V(z) + |\psi|^2 - \mu \right]\psi,
\label{gpe}
\end{eqnarray}
where the density $|\psi|^2$, length, time and energy are respectively measured in units of
$2a$, $a_{\perp} = \sqrt{\hbar/m \omega_{\perp}}$, $\omega_{\perp}^{-1}$ and $\hbar\omega_{\perp}$.
In the case of a harmonic trap, the external potential takes the form
$V(z)=(1/2)\Omega^{2} z^{2}$, where $\Omega = \omega_z/\omega_\perp \ll 1$ is the normalized
trap strength, which is a naturally occurring small parameter of the system.
The function $\gamma(z)$, which accounts for the dissipation, takes the form given in Eq.\eqref{eq:approx1}.

\begin{figure}[tbp]
\includegraphics[scale=0.275]{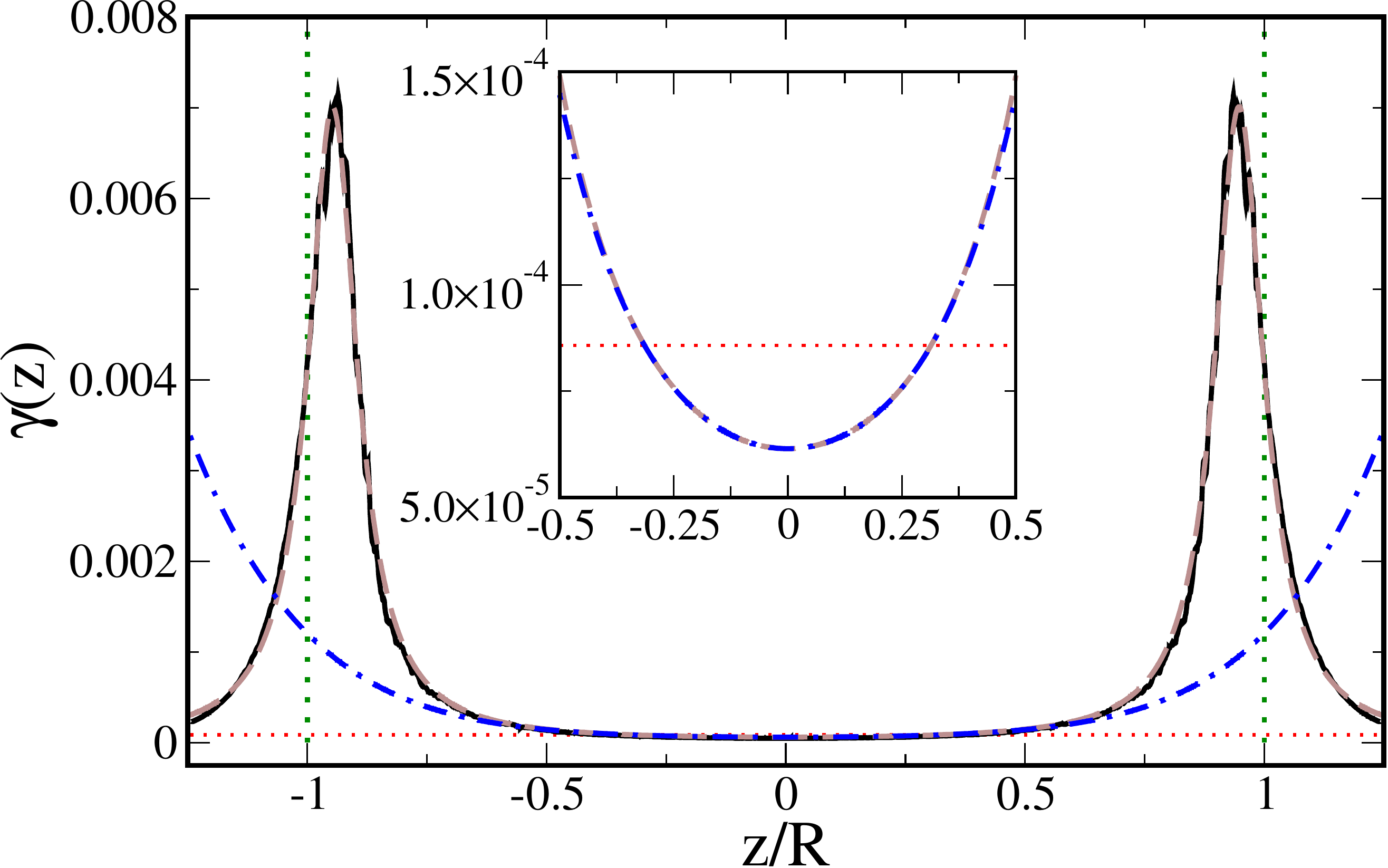}
\caption{(Color online) Shown is a comparison of $\gamma(z)$ as given by Eq.~(\ref{eq:gamma_int})
[solid black line], the fit given by Eq.~(\ref{eq:approx1}) [dashed brown line], and the approximation of Eq.~(\ref{g_tay}) [dot-dashed blue line]
for $T=150$ nK. The dotted (red) curve depicts the mean value of $\gamma(z)$, while the vertical dotted green lines indicate the points $z=\pm R$, which define the rims of the TF cloud. The inset shows a magnification of this figure for the spatial interval $[-R/2, R/2]$: there, the agreement between the fit to $\gamma(z)$ (Eq.~(\ref{eq:approx1}); dashed brown line) and the approximate form of this (Eq.~(\ref{g_tay}); dot-dashed blue line)
is excellent, as the corresponding curves essentially coincide and become indistinguishable.
}
\label{tayl_1}
\end{figure}
We now seek a solution of Eq. (\ref{gpe}) in the form
$\psi(z,t) = \psi_b (z,t)\exp[-i\theta(t)] \upsilon(z,t)$,
where $\psi_b(z,t)$ and $\theta(t)$ denote the background amplitude (which can be approximated in the framework of the Thomas-Fermi (TF) approximation) and phase respectively, while
the unknown complex function $\upsilon(z,t)$ will represent a dark soliton.
Assuming that the condensate dynamics involves a fast relaxation scale to the ground state, and that the dark soliton evolves on top of this ground state, we obtain the following perturbed nonlinear Schr\"{o}dinger equation for the dark soliton wave function \cite{Cockburn2010}:
\begin{equation}
i \partial_{t} \upsilon +\frac{1}{2}\partial_z^2 \upsilon -(|\upsilon |^{2}-1)\upsilon = P(\upsilon),
\label{pnls}
\end{equation}
where we have used the rescalings $t \rightarrow \mu t$ and $z \rightarrow \sqrt{\mu}z$. The total perturbation $P(\upsilon )$ in Eq.~(\ref{pnls}) has the form:
\begin{equation}
P(\upsilon )=\frac{1}{2\mu} \left[ 2 \left( 1-|\upsilon |^{2} \right)V(z)\upsilon
+ \frac{dV}{dz} \partial_z \upsilon + 2\mu \gamma(z) \partial_t \upsilon \right].
\label{pert}
\end{equation}

In the absence of the perturbation [$P(\upsilon)=0$], Eq.~(\ref{pnls}) is a conventional defocusing NLS equation,
which possesses a dark soliton solution of the form \cite{Zhakharov1973}:
\begin{equation}
\upsilon (z,t)=\cos \varphi \tanh Z +i \sin \varphi,
\label{ds}
\end{equation}
where $Z =\cos \varphi \left[ x-(\sin \varphi)t \right]$, and $\varphi$ is the ``soliton phase angle'' $(|\varphi|<\pi/2)$
describing the {\it darkness} of the soliton through the expression $|\upsilon|^{2}=1-\cos^{2} \varphi {\rm sech}^{2}Z$;
note that the limiting cases $\varphi=0$ and $\cos\varphi \ll 1$ correspond to the so-called {\it black} and {\it grey} solitons, respectively \cite{Kivshar1998,Frantzeskakis2010}.

The effect of perturbation (\ref{pert}) on the dark soliton will be treated analytically by means of the adiabatic approximation of the Hamiltonian approach of the perturbation theory for dark solitons. This approach was introduced in Ref.~\cite{Kivshar1994}
for the case of a constant background density, and subsequently used for trapped BECs in Ref.~\cite{Frantzeskakis2002} (see also the review  of Ref.~\cite{Frantzeskakis2010} and our previous work \cite{Cockburn2010}). According to this approach, the parameters of the dark soliton (\ref{ds}) become slowly-varying functions of time $t$, but the soliton's functional form remains unchanged. Thus, the soliton phase angle  becomes $\varphi \rightarrow\varphi(t)$ and, as a result, the soliton coordinate becomes $Z \rightarrow Z =\cos\varphi(t) \left[z-z_{0}(t) \right]$,  
where
\begin{equation}
z_{0}(t)= \int_{0}^{t}\sin\varphi(t^{\prime })dt^{\prime},
\label{cent}
\end{equation}
is the soliton center. The evolution of the parameter $\varphi$ is described by the following equation \cite{Kivshar1998,Frantzeskakis2010,Kivshar1994}:
\begin{equation}
\frac{d\varphi}{dt}=\frac{1}{2\cos^{2}\varphi \sin \varphi}
{\rm Re}\left\{\int_{-\infty}^{+\infty}P(\upsilon)\partial_t \upsilon^{\ast}dz
\right\}.
\label{phi}
\end{equation}
The integral in Eq.~(\ref{phi}) involves three terms [cf. Eq.~(\ref{pert})]: the first two terms (accounting for the hamiltonian part of the perturbation) can readily be evaluated upon expanding the potential $V(z)$ in Taylor series around the soliton's center, $z_0$; on the other hand, the third term (accounting for the dissipative part of the perturbation) can be evaluated by approximating the function $\gamma(z)$ by its Taylor expansion around the trap center ($z=0$). This expansion reads:
\begin{equation}
\gamma(z) \approx \gamma_0+\gamma_2 z^2+\gamma_4 z^4+ \gamma_6 z^6,
\label{g_tay}
\end{equation}
where the constant coefficients are given by:
\begin{eqnarray}
\gamma_0 &=& 2a/(c^2+d),\\
\nonumber
\gamma_2 &=& 2a(3c^2-d)/(c^2+d)^3,\\
\nonumber
\gamma_4 &=& 2a(5c^4-10c^2d+d^2)/(c^2+d)^5, \\
\nonumber
\gamma_6 &=& 2a(7c^6-35c^4d+21c^2d^2-d^3)/(c^2+d)^7.
\label{gamma_val}
\end{eqnarray}

Using, as an example, the values of $a$, $c$ and $d$ corresponding to $T=150$~nK (cf. Table~\ref{table_acd}), in Fig.~\ref{tayl_1} we compare the fit to $\gamma(z)$ given by Eq.~(\ref{eq:approx1}) with the approximate expansion of Eq.~(\ref{g_tay}). More specifically, we focus on the spatial interval of $[-R/2, R/2]$ (where $R$ is the TF radius of the cloud) -- see inset of Fig.~\ref{tayl_1}: this interval is particularly relevant for our analytical and numerical considerations (see below), as we are mainly concerned with solitons moving close to the trap center. As seen in the figure, in this regime, the approximate expression is almost identical to the more accurate one, thus justifying the degree of approximation used in Eq.~(\ref{g_tay}).

To this end, using the approximate expression of Eq.~(\ref{g_tay}), calculation of the integral in Eq.~(\ref{phi}) leads to the following result:
%
%
\begin{eqnarray}
\frac{d\varphi }{dt}=&-&\frac{1}{2}\cos \varphi \frac{dV}{dz} + \gamma_0 \frac{2}{3} \mu \cos \varphi \sin \varphi
\nonumber \\
&+& \gamma_2 \frac{\left(\pi^2-6\right)}{18} \mu \tan \varphi
\nonumber \\
&+& \gamma_4 \frac{\pi^2 \left(7\pi^2-60\right)}{360} \mu \tan \varphi {\rm sec^2} \varphi
\nonumber \\
&+& \gamma_6 \frac{\pi^4 \left(31\pi^2-294\right)}{2016} \mu \tan \varphi {\rm sec^4} \varphi.
\label{phif}
\end{eqnarray}
Next, combining Eqs.~(\ref{phif}) and (\ref{cent}), we obtain the following effective equation of motion for the
dark soliton center,
\begin{eqnarray}
\frac{d^{2}z_{0}}{dt^{2}} &=& \left[\gamma_0 \frac{2}{3} \mu  \frac{dz_{0}}{dt} - \left( \frac{\Omega}{\sqrt{2}}\right)^2 z_0 \right] \left[1 - \left(\frac{dz_{0}}{dt}\right)^2 \right]
\nonumber \\
&+& \frac{dz_{0}}{dt} \Bigg\{ \gamma_2 \left( \frac{ \pi^2-6}{18} \right) \mu
\nonumber \\
&+& \gamma_4 \left( \frac{ \pi^2 ( 7\pi^2-60)}{360} \right) \left[1 - \left(\frac{dz_{0}}{dt}\right)^2 \right]^{-1} \mu
\nonumber \\
&+& \gamma_6 \left( \frac{\pi^4 ( 31\pi^2-294) }{2016} \right) \left[1 - \left(\frac{dz_{0}}{dt}\right)^2 \right]^{-2} \mu
\Bigg\}.
\label{nl_em}
\end{eqnarray}
Notice that a variant of Eq.~(\ref{nl_em}), corresponding to $\gamma_2=\gamma_4=\gamma_6=0$, was presented in Ref.~\cite{Cockburn2010}, where the function $\gamma(z)$ was approximated by the constant value $\gamma_0$ (which is close to the mean value of $\gamma(z)$ in the interval $[-R/2,R/2]$ -- see Fig.~\ref{tayl_1}). Nevertheless, here we are going to analyze the more general problem and investigate the dissipative dynamics of solitons taking into regard the effect of the spatially-dependent profile of $\gamma(z)$. 

Here we should note that there appears to be a singularity in the solutions of Eq.~(\ref{nl_em}), corresponding to velocity values $dz_0/dt =1$, for which Eq.~(\ref{nl_em}) becomes invalid. This is also consistent with the analysis of Ref.~\cite{Pelinovsky2005}, where it is shown that formal perturbation theory fails for extremely shallow dark solitons (with phase angles $\varphi \approx \pi/2$). In any case, in the physically relevant scenarios that we consider in our simulations below, solitons are observed to decay at the rims of the condensate at times smaller than the one needed for the soliton velocity to become $dz_0/dt =1$. 

\subsection{The equation of motion for the soliton center}

The equation of motion (\ref{nl_em}) is evidently nonlinear. Nevertheless, assuming that the dark soliton is close to a black one (i.e., $\varphi$ is sufficiently small), Eq.~(\ref{phif}) can be reduced to the following linearized form,
\begin{equation}
\frac{d^{2}z_{0}}{dt^{2}} - \tilde{\gamma_0} \mu \frac{dz_{0}}{dt} + \left( \frac{\Omega}{\sqrt{2}}\right)^2 z_0 = 0,
\label{em}
\end{equation}
where we have introduced the variable
\begin{eqnarray}
\tilde{\gamma_0} &=& \frac{2}{3} \gamma_0 + \frac{\left(\pi^2-6\right)}{18} \gamma_2
\nonumber \\
&+& \frac{\pi^2 \left(7\pi^2 -60\right) }{360} \gamma_4 + \frac{\pi^4 \left(31\pi^2-294\right)}{2016} \gamma_6.
\label{nl_em_2}
\end{eqnarray}
Equation (\ref{em}) is similar to the equation of motion derived in Ref.~\cite{Fedichev1999} by means of a kinetic-equation approach. In the limiting case of zero temperature (i.e., for $\tilde{\gamma_0}=0$), Eq. (\ref{em}) recovers
the well-known result (see, e.g., Refs.~\cite{Frantzeskakis2002,Fedichev1999,Busch2000b,Parker2003,Konotop2004,Pelinovsky2005,Theocharis2005}) 
that a dark soliton oscillates with constant amplitude and frequency $\Omega/\sqrt{2}$ in the harmonic trap $V(z)=(1/2)\Omega^{2} z^{2}$. On the other hand, at finite temperatures (i.e., for $\tilde{\gamma_0} \ne 0$), the linearized equation of motion (\ref{em}) additionally incorporates an anti-damping term, $\propto -dz_0/dt$ [with a coefficient that takes into account the spatial dependence of $\gamma(z)$], which describes the expulsion of the dark soliton due to the interaction with the thermal cloud.

It is clear that the nature of the solutions of Eq.~(\ref{em}) depend on whether the roots of the auxiliary equation $s^2 -(2/3)\tilde{\gamma_0}\mu s + (\Omega/\sqrt{2})^2 = 0$ are real or complex. The roots are given as
\begin{equation}
s_{1, 2} = \frac{1}{3}\tilde{\gamma_0} \mu \pm \left(\frac{\Omega}{\sqrt{2}}\right) \sqrt{\Delta},
\,\,\,\,\,
\Delta =  \left( \frac{\tilde{\gamma_0}}{\gamma_{cr}} \right)^2 -1,
\label{s12}
\end{equation}
where $\gamma_{cr}=(3/\mu)(\Omega/\sqrt{2})$, and the discriminant $\Delta$ determines the type of the motion:\\
In the {\it super-critical} case of strong anti-damping with $\Delta>0$,
i.e., for high temperatures such that $\tilde{\gamma_0} > \gamma_{cr}$,
the soliton trajectory is given by
\begin{equation}
	\begin{split}
		z_0(t)= &\frac{1}{s_2-s_1}[(s_2z_0(0)-\dot{z}_0(0)) \exp(s_1 t)\\
		&+(\dot{z}_0(0)-s_1 z_0(0))\exp(s_2 t)].
	\end{split}
\label{spcr}
\end{equation}
In the {\it critical} case with $\Delta=0$, i.e., $\tilde{\gamma_0} = \gamma_{cr}$,
the soliton trajectory is given as
\begin{equation}
	z_0(t)=[z_0(0) + (\dot{z}_0(0)
	-\frac{1}{3}\tilde{\gamma}_0 \mu z_0(0))t]
	\exp(\frac{1}{3}\tilde{\gamma}_0 \mu t).
\label{cr}
\end{equation}
Finally, in the {\it sub-critical} case of weak anti-damping with $\Delta<0$, i.e., for sufficiently low
temperatures such that $\tilde{\gamma_0} < \gamma_{cr}$, the soliton trajectory is given by
\begin{equation}
	\begin{split}
		z_0(t) = &\exp(\frac{1}{3}\tilde{\gamma}_0\mu t)
		[z_0(0)\cos(\omega_{\rm osc}t)\\
		&+ \omega_{\rm osc}^{-1} (\dot{z}_0(0)-\tilde{\gamma}_0\mu
		z_0(0))\sin(\omega_{\rm osc}t) ],
	\end{split}
	\label{sbcr}
\end{equation}
where
\begin{equation}
\omega_{\rm osc} = \left(\frac{\Omega}{\sqrt{2}}\right) \left[1-\left(\frac{\sqrt{2}}{3} \frac{\tilde{\gamma_0} \mu}{\Omega}\right)^2 \right]
\label{osc}
\end{equation}
is the soliton oscillation frequency.
\begin{figure}[tbp]
	\includegraphics[scale=0.3]{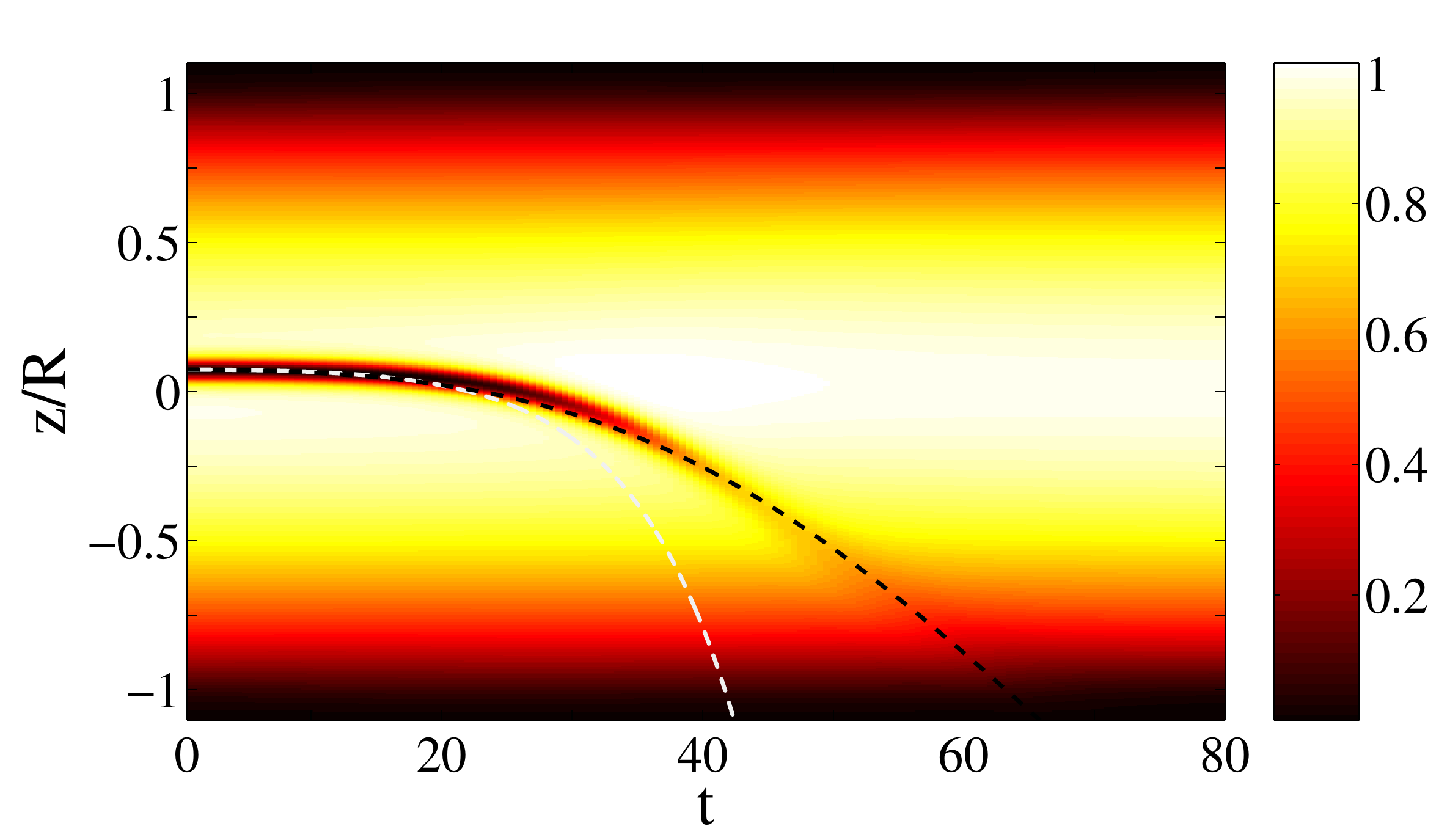}
	\includegraphics[scale=0.3]{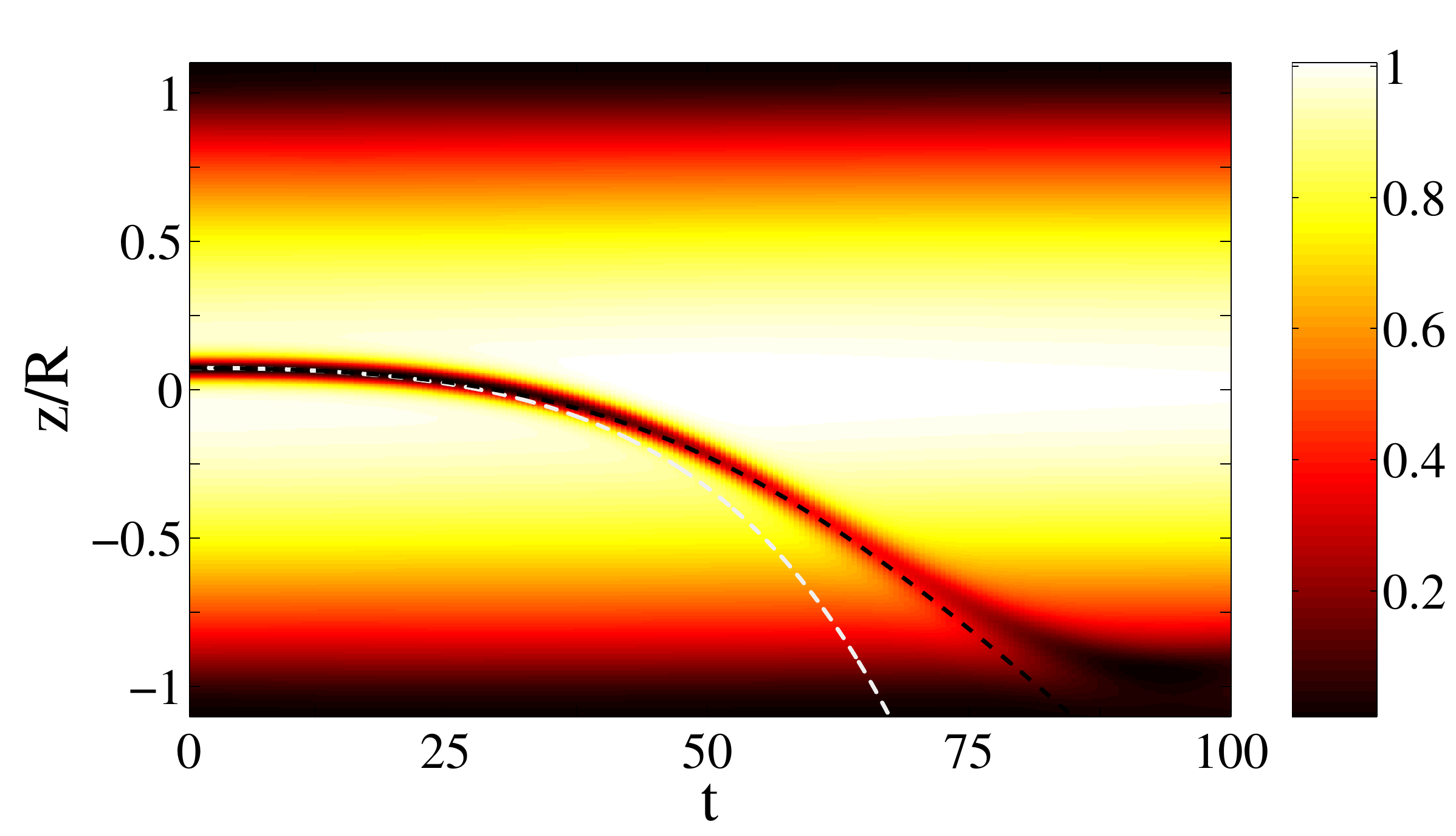}
	\includegraphics[scale=0.3]{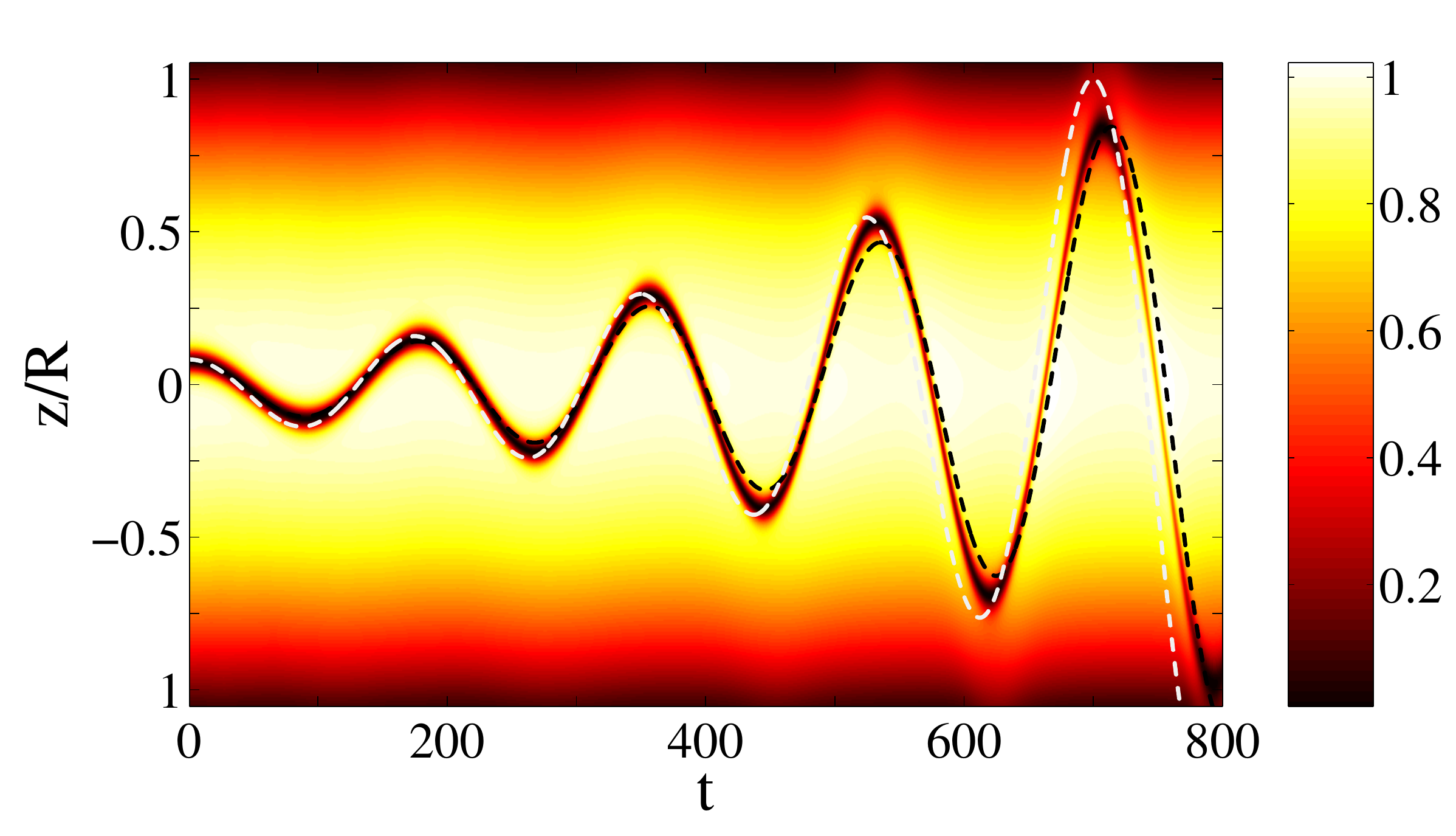}
	\caption{(Color online) Spatiotemporal evolution of the density of a BEC
	confined in an harmonic trap with $\Omega=0.05$ and $\mu=1$,
	with a dark (black) soliton initally placed at $z_0=2$
	for the super-critical case with $\tilde{\gamma_0}/\gamma_{cr}= 1.89$  
	(top panel; 
	$a=84.44$, $c=21.67$ and $d=93.97$),
	critical case with $\tilde{\gamma_0}/\gamma_{cr}=1$ 
	(middle panel; $a=43.67$, $c=22.19$ and $d=59.16$),
	and sub-critical case with $\tilde{\gamma_0}/\gamma_{cr}= 0.09$ 
	(bottom panel; $a=4.53$, $c=23.89$ and $d=19.72$). Here $\gamma_{cr}= 0.106$.
	The white dashed lines across the soliton trajectories correspond to the analytical prediction of
	the linearized equation of motion (\ref{em}), while the black dashed lines are obtained by solving numerically the 		nonlinear equation of motion (\ref{nl_em}).
	}
\label{fig2}
\end{figure}

The above simple analysis shows that in the case of relatively high-temperatures (with
$\tilde{\gamma_0} \ge \gamma_{cr}$), the dark soliton will not `survive' long enough to oscillate in the trap, a result being in
qualitative agreement with experimental observations of dark matter-wave solitons in elongated Bose gases
\cite{Burger1999}.
On the other hand, when the temperature is relatively low (i.e., $\tilde{\gamma_0} < \gamma_{cr}$),
the dark soliton performs oscillations with an increasing amplitude and period -- recall that the oscillation frequency is down-shifted as per Eq.~(\ref{osc}) with respect to its value $\Omega/\sqrt{2}$, at zero-temperature. These analytical results are in qualitative agreement with the numerical ones obtained recently in the framework of the Zaremba-Nikuni-Griffin model \cite{Jackson2007,Jackson2007b}.

\subsection{Non-linear vs. linearized equation of motion}

To assess the above model, we now compare analytical 
[Eqs.~(\ref{nl_em}) - (\ref{nl_em_2})] and numerical [Eqs.~(\ref{gpe}) with (\ref{eq:approx1})] soliton trajectories,
in each case making use of the approximate form for $\gamma(z)$ employed in the analytical approach.
First, we consider Eq.~(\ref{gpe}), with $\gamma(z)$ given in Eq.~(\ref{g_tay}), with an initial condition corresponding to a dark (black) soliton, initially placed off-centre at $z_0(0)=2$ (with zero initial velocity, $dz_0(0)/dt=0$), on top of a TF cloud, with $\mu =1$, confined in a trap of strength $\Omega = 0.05$. The resulting soliton trajectories, found by integrating Eq.~(\ref{gpe}) by means of the split-step Fourier method, are shown in Fig.~\ref{fig2} 
for the
super-critical ($\tilde{\gamma}/\gamma_{cr}=1.89$, top plot), 
critical ($\tilde{\gamma}=\gamma_{cr}=0.106$, middle) 
and sub-critical ($\tilde{\gamma}/\gamma_{cr}= 0.09$, bottom) 
cases.
The black dashed lines in each plot depict the numerically obtained solutions of Eq.~(\ref{nl_em}), while the white dashed lines show the corresponding analytical solutions of Eq.~(\ref{em}).

Generally, it is observed that the results based on our analytical approximations, i.e., the solutions of Eqs.~(\ref{nl_em}) and (\ref{em}) -- which were derived by employing the approximate form of $\gamma(z)$ [cf. Eq.~(\ref{tayl_1})] -- are in good agreement with the numerical results obtained by the DGPE with the exact form of $\gamma(z)$ [cf. Eq.~(\ref{g_tay})]. Nevertheless, the solutions of the nonlinear equation of motion (\ref{nl_em}) are more accurate than the analytical solutions of Eq.~(\ref{em}) in capturing the soliton trajectories obtained by the DGPE. This behavior is more pronounced for longer times, where the soliton either decays (top and middle panels of Fig.~\ref{fig2}) or performs large amplitude oscillations (bottom panel of Fig.~\ref{fig2}): in fact, the solutions of the nonlinear equation of motion are able to correctly predict the decay time of the solitons [which is underestimated by the solutions of Eq.~(\ref{em})] in the super-critical and critical cases of strong dissipation, or follow quite accurately the DGPE trajectory in the sub-critical case of weak dissipation; notice that, in the latter case, the analytical solution of Eq.~(\ref{em}) underestimates (overestimates) the frequency (amplitude) of oscillation for longer times. 

We now proceed with a systematic comparison between analytical approximations,
focussing on the more accurate nonlinear equation of motion \eqref{nl_em}, 
and numerical (DGPE and SGPE) results. 

\subsection{Comparison between numerical (SGPE and DGPE) results and analytical approximations}

To ensure long soliton lifetimes for this comparison, 
we focus on one of the lower temperatures within our study,
$T=150$~nK, which nevertheless still corresponds to $k_{B}T=0.8\mu$. 
We 
perform a direct numerical integration of the DGPE, Eq.~(\ref{gpe}) 
[with $\gamma(z)$ given by Eq.~(\ref{eq:approx1})]
and compare this to respective results obtained via the analytic (\ref{nl_em}) and SGPE models. 
The initial condition takes the form of a dark soliton, initially placed at the trap centre $z_0(0)=0$ with initial 
velocity $dz_0(0)/dt=0.25$ (the other parameters also remain the same with 
$\mu =1.58\hbar\omega_{\perp}$ and the dimensionless trap strength $\Omega = 4\times 10^{-3}$). 

\begin{figure}[tbp]
\includegraphics[width=9cm,height=6cm]{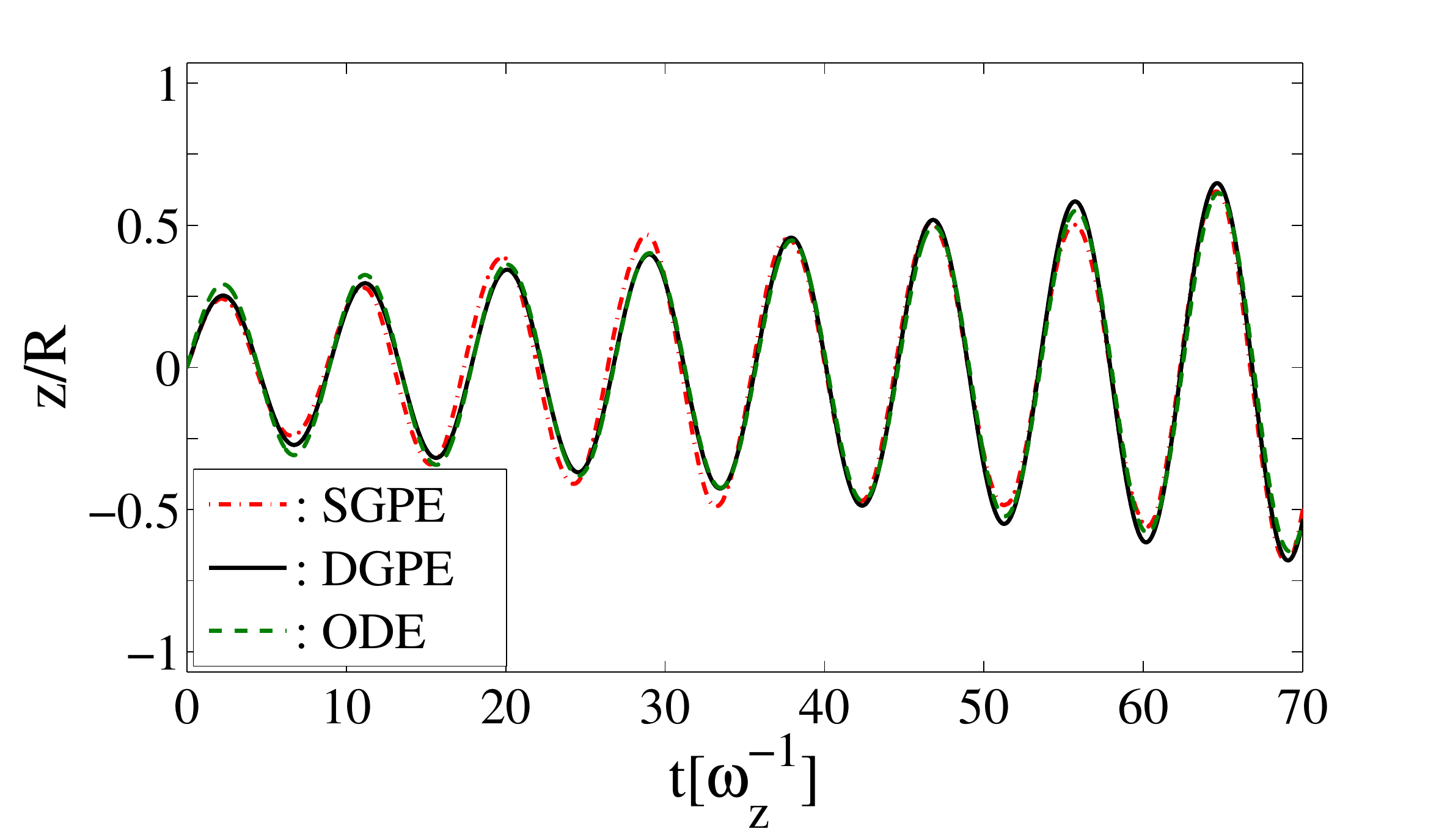}
\caption{(Color online) Soliton trajectories as found by the numerical integration of the DGPE [solid (black) line], the SGPE [dash-dotted (red) line; single run with a decay time from the mean bin], 
and the nonlinear ordinary differential equation (ODE), Eq.~(\ref{nl_em}) [dashed (green) line], for $T=150$~nK.
Here $a=0.0219$, $c=26.63$, and $d=3.128$.
}
\label{new_fig2}
\end{figure}

In Fig.~\ref{new_fig2}, we show the soliton trajectories found via the DGPE and the SGPE, as well the solutions of the nonlinear equation of motion (\ref{nl_em}) stemming from our analytical approximations. It is clear that not only the solution of the DGPE captures quite accurately the one of the SGPE (similarly to the behaviour found in Ref.~\cite{Cockburn2010}), but also the result of the analytical approximation is in very good agreement with the numerical results of the DGPE and SGPE. The SGPE trajectory shown is from a single run with a decay time close to
the ensemble average; for these parameters, such trajectories were found to display dynamics close
to the ensemble average in Ref.~\cite{Cockburn2010}.

A further comparison between the analytical model above and the SGPE/DGPE can be seen in
Fig.~\ref{fig:time_analysis}, where the predicted average decay times are plotted.
In order to make this comparison, the decay times in the analytical model must be
extracted based on the visibility of the soliton over the background thermal 
fluctuations, as we discuss in the following Section.

\section{Soliton visualization}
\label{sec:soli_vis}

\subsection{Single shot fluctuation issues}

\begin{figure}[tbp]
\includegraphics[scale=0.34,clip]{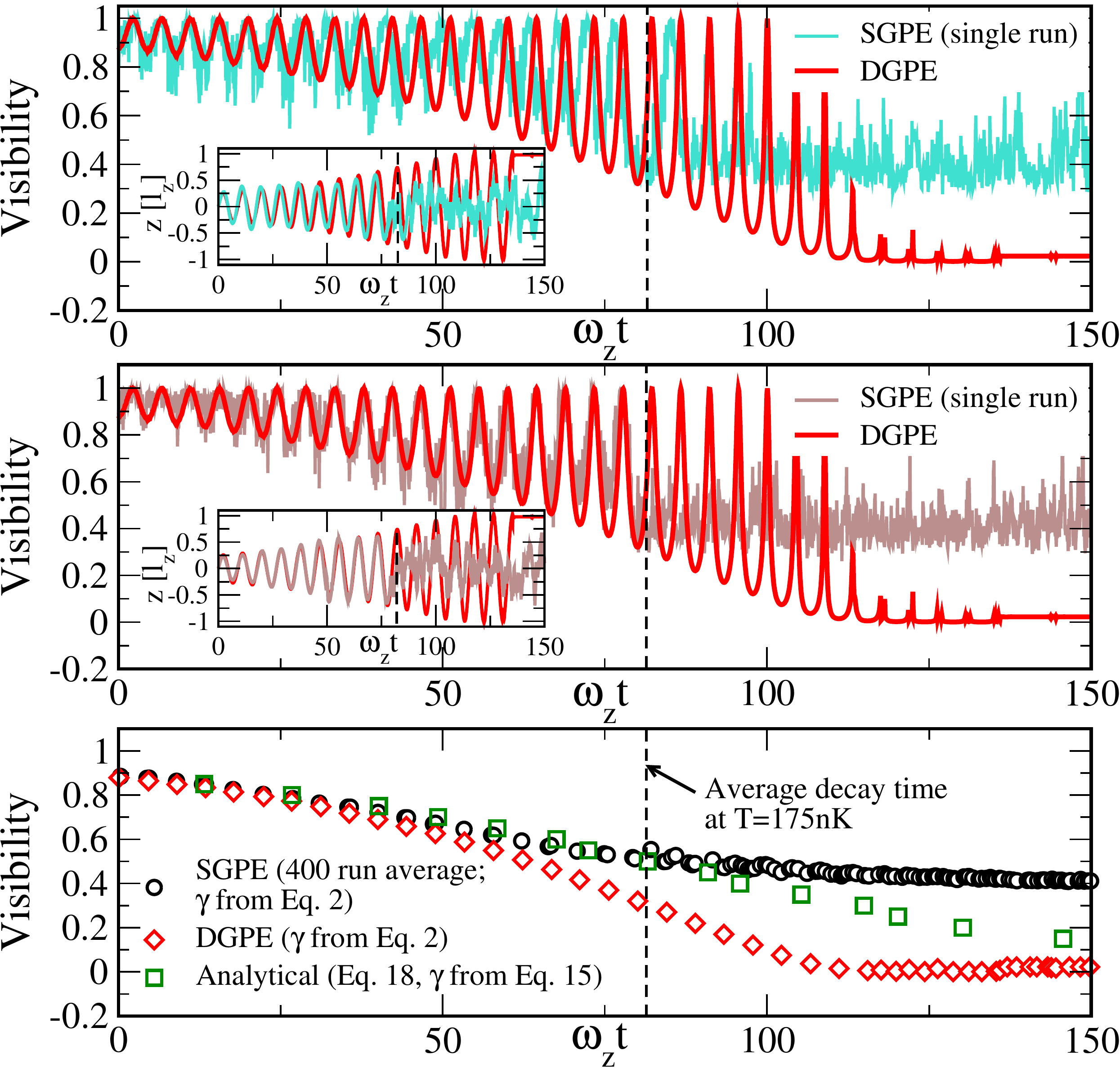}
\caption{(Color online) Numerical results for the visibility as a function of time for two example SGPE 
					single runs (noisy turquoise, top plot; noisy brown, middle plot) 
					vs. the DGPE (smooth red curve in top and middle plots); the insets 
					show the corresponding soliton trajectories. 
					The lower plot shows the analytical result [green squares] versus the {\it average} visibility 
					from the SGPE (black circles)
					and the visibility from the DGPE (red diamonds; 
          the latter two data sets correspond to the local minima of oscillations, 
					which can be seen in the upper plots. 
					At this	temperature ($T=175$nK), the analytical prediction is very good up to the point that
					the soliton can be tracked, though uses the approximate form for
					$\gamma(z)$ of Eq.~\eqref{g_tay} while the SGPE and DGPE use Eq.\eqref{eq:gamma_int}. 
					The stochastic results are examples of
					solitons with decay times close to the ensemble mean, indicated by the vertical
					dashed line in all plots.}
\label{fig:Vis_fig}
\end{figure}

A quantity of relevance to experiments is the so-called visibility of the soliton. 
This is defined as \cite{Scott1998}:
\begin{equation}
\mathcal{V} =\frac{n_{\rm max}-n_{\rm min}}{n_{\rm max}+n_{\rm min}},
\label{visi}
\end{equation}
where $n_{\rm max}$ is the maximum BEC density and $n_{\rm min}$ is the minimum one, as set by the presence of the dark soliton. 
This parameter is a measure of how clearly a soliton can be seen in experiments. The SGPE accounts for both dissipation and fluctuations and can therefore be expected to produce experimentally-realistic
visibility predictions; two examples of the visibility of solitons produced 
by the SGPE are shown in Fig.~\ref{fig:Vis_fig} (top and middle plots),
showing an oscillatory decrease up to the point (denoted by the vertical dashed lines) when the soliton can no longer be distinguished from the background density fluctuations.

In the displayed example trajectories, the solitons are lost to the background when the visibility decreases below around $50\%$.
For comparison, we note that in the Hannover experiment \cite{Burger1999}, they reported a contrast
in the range $20\%-40\%$ throughout their measurements, and 
a visibilty of $50\%$ in our present work corresponds to a contrast of $n_{\rm min}/n_{\rm max}\sim 33\%$.
The corresponding DGPE prediction is also shown (red curves) revealing good
agreement in the region where the soliton can actually be monitored over 
fluctuations in the stochastic cases (top and middle plots). 
Clearly the results become meaningless beyond that time, 
as the soliton is then lost within that stochastic realization;
betind this time the DGPE soliton signal remains visible, but this is due
to incorrectly neglecting fluctuations. 

The periodic behaviour in the DGPE and SGPE visibility arises 
because the soliton depth oscillates as the soliton traverses the harmonic trap.
The points at which $\mathcal{V}=1$ correspond to the turning points of the 
soliton motion, when the soliton depth is equal to the background density 
(or equivalently $n_{\rm min}=0$).
Perhaps a better measure is to look at the evolution at a specific point in the trap, e.g. the trap centre, which corresponds to the minimum visibility during the 
dissipative soliton motion. This is shown in Fig.~\ref{fig:Vis_fig} (bottom) and reveals
quite good agreement between DGPE and SGPE, up to around the time that, on average,
the SGPE solitons were visible.

We now want to consider the corresponding analytical prediction based on our model of Section~\ref{sec:analytics}.
As when comparing the numerical DGPE results to those of the SGPE (see Fig.~\ref{fig:time_analysis} 
of Section \ref{sec:distr}), we again define the condition for soliton decay 
based upon the level of background thermal fluctuations obtained from
the SGPE. Then, calculating an expression for the soliton visibility analytically, we are able to 
predict analytically the lifetime of a soliton within a finite temperature gas,
accounting for both dissipation and background fluctuations.
Since $n_{\rm min} \sim \mu-\mu \cos^2 \varphi$ (recall that the soliton depth is $\sqrt{\mu}\cos \varphi$), the visibility can be expressed in terms of the soliton's phase angle as
\begin{equation}
\mathcal{V}=\frac{\cos^2 \varphi}{1+\sin^2 \varphi}.
\label{visi2}
\end{equation}
Thus, $0 \le \mathcal{V} \le 1$, with the limiting values $\mathcal{V}=0$ and $\mathcal{V}=1$ corresponding, respectively, to a shallow soliton with $\varphi \rightarrow \pi/2$ and a stationary kink with $\varphi \rightarrow 0$. Apparently, since $\varphi=\varphi(t)$, the visibility is generally a function of time, but its analytical form can be determined via the time-dependence of $\varphi(t)$, which can be derived numerically by means of Eq.~(\ref{phif}) (see Section~\ref{sec:an_approx}). Nevertheless, it should be noticed that a simple analytical expression for the visibility can also be obtained in the case of sufficiently deep solitons ($\cos \varphi \approx 1$ and  $\sin \varphi \approx \phi$) oscillating in a small region around the trap center [i.e., for $V(z=0)=0$]: in this case, Eq.~(\ref{phif}) becomes $d\varphi/dt = \gamma_{\rm o} \varphi$, and leads to the result
$\varphi(t)=\varphi_{\rm o}(t) \equiv \varphi(0) \exp(\gamma_{\rm o} \mu t)]$
(here, $\varphi_0$ is the initial value of the phase angle) and, accordingly, $\mathcal{V}|_{z=0} =\cos^2 \varphi_{\rm o}(t)/[1+\sin^2 \varphi_{\rm o}(t)]$.

Following the above arguments, we may
estimate the soliton lifetime and
the relevant results of the semi-analytical approximation, Eq.~(\ref{visi2}), are shown in
Fig.~\ref{fig:time_analysis} of Section \ref{sec:distr}.

We can also calculate the soliton visibility versus time analytically,
an example of which is shown in Figure~\ref{fig:Vis_fig}(c), alongside the
numerical DGPE and {\it average} SGPE results at the same temperature. 
The SGPE results in this case are an average over the visibility from few hundred runs,
which smears out the oscillatory behaviour, but yields a bulk behaviour consistent
with that of the single realizations above.
The agreement is very good between all approaches during the period that the soliton 
is visible over the background noise,
beyond which time (vertical dashed lines), the SGPE visibility (black circles)
plateaus.
Here we have chosen to compare the analytical data to the
{\it minima} of both the DGPE and {\it average} SGPE oscillations.
A departure of the numerical DGPE results from the SGPE data
occurs close to the average soliton decay time, at which point
many of the SGPE solitons have decayed, leaving a visibility reading 
in many single runs which corresponds to the background noise signal.
The DGPE soliton signal instead persists for much lower visibility values
as the physical effects of fluctuations are not included,
as also evident from the comparison of stochastic and dissipative 
`carpet plots' of Fig.~\ref{fig:front}.
The analytical results are based on the Taylor expansion for $\gamma(z)$
given by Eq.~\eqref{g_tay}, which for large $z$ is somewhat smaller
than the full form of $\gamma(z)$ used in the numerical DGPE and SGPE simulations
(see Fig.~\ref{tayl_1}).
This difference becomes apparent for times close to and after the average decay time 
(vertical dashed line), for which the numerical and analytical results deviate
slightly for this reason.

Nonetheless, based on the level of background noise within the SGPE, and the analytical 
formula for the visibility, it should be possible to predict realistic lifetimes for solitons 
within experiments based on this approach.

\subsection{Comments on related work}

The study of dark solitons using stochastic and classical field methods has received
significant attention recently 
\cite{Negretti2008,Cockburn_PhD,Damski2010,Cockburn2010,Martin2010a,Martin2010b}.
In particular, the SGPE was applied initially in parallel, independent works 
\cite{Damski2010,Cockburn2010}, and recently also in \cite{KWright2011}.
The analysis presented in \cite{Damski2010} considered dark solitons as 
relics due to a quench of the system parameters within a homogeneous, periodic
1D Bose gas. Notably, the distribution reported in \cite{Damski2010} 
describing the number of solitons 
with time (Fig.3 of Ref.~\cite{Damski2010}) has a 
form which is qualitatively similar to the decay time
distributions of Refs.~\cite{Cockburn2010,Cockburn_PhD} and 
in the present work.

More recently, Wright and Bradley \cite{KWright2011} 
performed a study which is more closely related to Ref.~\cite{Cockburn2010}
and our current analysis,
based on the stochastic projected Gross-Pitaevskii
equation (SPGPE) \cite{Gardiner2003,Blakie2008}. 
(This method is a variant of the SGPE \cite{Proukakis2008} which
features a projector into low energy modes.)
They found that the stochastic simulations yielded average
velocities which were lower than those found within dissipative 
GPE simulations,
implying that the solitons have longer lifetimes, on
average, within the stochastic simulations, i.e., that the noise
prolongs the solitons' existence.
This is in agreement with the long tails in the decay time
distributions of both Refs.~\cite{Cockburn_PhD,Cockburn2010} and the current work. 
Furthermore, the analytical
findings of Ref.~\cite{KWright2011}, and in particular Eq.(18) of Ref.~\cite{KWright2011},
arise as a special case of our previously reported results \cite{Cockburn2010} (see sentence preceeding Eq. (6) in \cite{Cockburn2010}) and by extension also of this work, which additionally treats a spatially-dependent dissipative term.
This can be seen from Eq.~\eqref{phif} above by setting $V=0$ (for a homogeneous system) and replacing $\gamma(z) \rightarrow \gamma_0$ (i.e. $\gamma_2 = \gamma_4 = \gamma_6 = 0$ for a spatially-constant dissipation), and multiplying through by $\cos(\phi)$~\footnote{In the 
notation of Ref.~\cite{KWright2011}, the velocity $\nu \equiv \sin(\phi)$, and 
hence, upon making these simplifications and coordinate transformation, Eq.(18) of 
Ref.~\cite{KWright2011} takes the form 
$d\phi/dt = (2/3) \gamma v \sqrt{1-v^2}$, which agrees with
Eq.~\eqref{phif} of Section \ref{sec:an_approx} 
(and also the in-text equation preceeding Eq. (6) in \cite{Cockburn2010})
in this limiting regime.
In comparing the above, note also our re-scaling of time $t\rightarrow \mu t$,
as we indicate below Eq.~\eqref{pnls} of Section~\ref{sec:an_approx}}.

Beyond this, it is not straightforward to 
give a more direct comparison to their findings
for several reasons:
(i) we consider a harmonically trapped system, while they focussed
on soliton propagation within a homogeneous sample; 
(ii) they measure the increase in velocity as the soliton decays,
which is straightforward without a trap,
while in our case the soliton velocity is {\it constantly changing} 
as it oscillates; an upshot of this is that the soliton
depth does not change monotonically as it decays, but has an
additional oscillatory component, complicating the velocity analysis;
(iii) we analyze the ensemble of solitons via their decay times,
and therefore effectively sample members of the SGPE ensemble at different times,
while they consider \emph{equal time} measurements of the
velocity via an ensemble measurement at a particular time.

\section{Conclusions}

We have characterized the dynamics of dark solitons propagating within
a partially condensed, harmonically trapped Bose gas in the presence of
phase fluctuations. Our analysis was based on 
a stochastic Gross-Pitaevskii equation, which reduces to a dissipative 
Gross-Pitaevskii equation upon neglect of the additive noise term.

Stochastic simulations allowed us to perform a statistical analysis
on the soliton decay times. 
Our results showed that to study soliton decay, information 
should be extracted from single soliton realizations prior to performing such an analysis,
in agreement with related experimental findings \cite{Weller_thesis}.
On doing so, we found dark soliton lifetimes
to be approximately lognormally distributed, which implies that some solitons
within a number of realizations may have very long lifetimes relative to the 
ensemble mean, an effect already observed in experiments \cite{Becker2008}.
We found the standard deviation and skewness of these distributions
to increase monotonically, and approximately linearly with temperature.
Extracting expectation values from the decay time distribution
obtained at each temperature, 
showed the purely dissipative results matched these stochastic expectation 
values decay times well, once the effects of background fluctuations
were taken into account.

Considering the interplay between noise in the initial conditions
(used e.g. in some simpler approximate models) and the dynamical noise at each temporal step of the Stochastic Gross-Pitaevskii equation - which loosely corresponds to a stochastic random  kick to the soliton position -, 
we found the dynamical noise to play an important role in determining the final decay time.

We also presented results for the experimentally relevant visibility of the soliton and found the soliton decay to be related to the visibility in our numerical simulations reaching a plateau (at which point our soliton tracking algorithm breaks down). The value of the plateau is set by the strength of the background fluctuations, which is a direct measure of temperature in these systems.

In the purely dissipative case, we derived analytical expressions
describing the dynamics of the soliton density notch,
with good agreement found between these and numerical
solution to the dissipative Gross-Pitaevskii equation, generalizing our
earlier work \cite{Cockburn2010} to the case of a spatially
dependent damping, as obtained
{\it ab initio} within the stochastic Gross-Pitaevskii formalism.
The average soliton decay times were found to scale
as $T^{-4}$; this has been previously obtained for homogeneous systems
at low temperatures $k_{B}T\ll\mu$, but our numerics indicates that
(at least within the classical field approximation for the low-lying modes of
the system) this can be extended to the trapped case, and to the regime for which 
$k_{B}T\lesssim\mu$, even in the presence of phase fluctuations.

Observing the dynamics of dark solitons within phase-fluctuating
condensates offers an intriguing opportunity to observe a macroscopic
quantum object undergoing a Brownian-like motion. We hope that the
fluctuating aspects of this behavior, and
especially the dependence on temperature,
may be further analyzed in
future experiments on ultracold gases in the near future.

\section*{Acknowledgments} 
We gratefully acknowledge funding from EPSRC (SPC, NPP), NSF-DMS-0806762 and from the Alexander von Humboldt Foundation (PGK), 
and the Special Account for Research Grants of the University of Athens (DJF). 
We thank Kai Bongs, Christian Gross and Markus Oberthaler for useful discussions.

\bibliographystyle{apsrev4-1}
\bibliography{references}

\appendix

\end{document}